\documentclass[%
 aip,
 amsmath,amssymb,
reprint,%
]{revtex4-1}

\usepackage{graphicx}
\usepackage{dcolumn}
\usepackage{bm}

\usepackage[utf8]{inputenc}
\usepackage[T1]{fontenc}
\usepackage{mathptmx}
\usepackage{etoolbox}
\usepackage{hyperref} 
\usepackage{caption}

\makeatletter
\def\@email#1#2{%
 \endgroup
 \patchcmd{\titleblock@produce}
  {\frontmatter@RRAPformat}
  {\frontmatter@RRAPformat{\produce@RRAP{*#1\href{mailto:#2}{#2}}}\frontmatter@RRAPformat}
  {}{}
}%
\makeatother

\begin{document}


\title{Revealing recurrent regimes of mid-latitude atmospheric variability using
novel machine learning method}

\author{Dmitry Mukhin}
\email[Dmitry Mukhin: ]{mukhin@ipfran.ru}
\affiliation{Institute of Applied Physics of the Russian Academy of Science, Nizhny Novgorod, Russia}
\author{Abdel Hannachi}
\affiliation{Department of Meteorology, Stockholm University, Stockholm, Sweden}
\author{Tobias Braun} 
\affiliation{Potsdam Institute for Climate Impact Research, Potsdam, Germany}
\author{Norbert Marwan}
\affiliation{Potsdam Institute for Climate Impact Research, Potsdam, Germany}


\date{\today}

\begin{abstract}
The low frequency variability of the extratropical atmosphere involves hemispheric-scale recurring, often persistent, states known as teleconnection patterns or regimes, which can have profound impact on predictability on intra-seasonal and longer timescales. However, reliable data-driven identification and dynamical representation of  such states are still challenging problems in modeling dynamics of the atmosphere.       
We present a new method, which allows both to detect recurring regimes of atmospheric variability, and to obtain dynamical variables serving as an embedding for these regimes. The method combines two approaches from nonlinear data analysis:  partitioning a network of recurrent states with studying its properties by the recurrence quantification analysis and the kernel principal component analysis. We apply the method to study teleconnection patterns in a quasi-geostrophical model of atmospheric circulation over the extratropical hemisphere 
as well as to reanalysis data of geopotential height anomalies in the mid-latitudes of the Northern Hemisphere atmosphere in the winter seasons from 1981 to the present. It is shown that the detected regimes as well as the obtained set of dynamical variables explain large-scale weather patterns, which are associated, in particular, with severe winters over Eurasia and North America. The method presented opens prospects for improving empirical modeling and long-term forecasting of large-scale atmospheric circulation regimes.
\end{abstract}

\maketitle

\begin{quotation}
Behavior of weather systems over the mid-latitudes is well-known as strongly chaotic and having a very limited horizon of reliable forecasting. While movements of synoptic-scale structures like cyclones and anticyclones are predicted well within 1-2 weeks, larger structures of atmospheric circulation  with longer time  scales are still poorly investigated. As it is shown by models and data analysis, dynamics on these time scales, also called the low-frequency variability, is characterized by recurrent global patterns, or regimes, which can strongly impact long-term weather conditions in different regions. However, both identification and dynamical representation of such regimes based on data is a controversial problem due to the lack of robust and reliable methods of data analysis/data analysis methods. Here we suggest a method which allows to detect the regimes and, simultaneously,
to obtain dynamical variables representing their dynamics. The
method involves and joins together/combines several approaches from nonlinear data analysis: partitioning
a network of recurrent states, recurrence
quantification analysis and nonlinear principal component analysis.  Studying winter low frequency variability (LFV) 
in the Northern Hemisphere mid-latitudes by the suggested method allows us to reveal and investigate dynamical properties of large-scale weather patterns, which are associated, in particular, with severe winters over Eurasia and North America. The results presented open prospects for improving data-driven modeling and long-term forecasting of large-scale atmosphere circulation regimes.
\end{quotation}

\section{Introduction}

Many non-linear multidimensional systems exhibit chaotic behavior with a continuum of time scales, are poorly predictable, and are generally difficult to distinguish from a random process. However, in the state space of the system there may be sets of states in which the system is found more often than others. Such intermittently recurrent states can have varying lifetimes (or persistence) and regularity of occurrence. Their study is important from a practical point of view, because in the space of observables they correspond to the most typical regimes of the system dynamics. However, both the identification and analysis of their dynamical properties based on the observations are still challenging, especially when the observations live in a high-dimensional space, such as our weather and climate system \cite{Hannachi21}.

Studies of atmospheric variability often distinguish between synoptic scales which embed day-to-day
variability
steered by baroclinic instability in the storm track region, and low-frequency 
variability with longer time scales.
Compared to baroclinic processes, low-frequency variability is still not entirely well understood and is known to be challenging to model and predict \cite{SuraHannachi2015}.
 Low-frequency variability embeds, in particular, tropospheric planetary waves and coherent
 large-scale structures including blocking and different phases of teleconnections, such as the
 North Atlantic Oscillation (NAO).
 A proper understanding of low-frequency variability proves invaluable in improving weather/climate
 prediction on intraseasonal and/or seasonal timescales, in addition to many other applications not
 least subgrid parametrization, climate change feedback, downscaling etc, see, e.g., the review 
 \onlinecite{Hannachi+2017} and references therein.
 
The atmospheric system is a highly nonlinear dynamical system 
with complex 
interactions between very many degrees-of-freedom involving different temporal and spatial 
scales.
As far as large scale flow is concerned there is evidence of the existence of preferred recurrent 
 and persistent circulation patterns \cite{Hannachi+2017}. For example, it is well known that
 northern hemisphere (NH) low-frequency variability is partly manifested by several teleconnection
 patterns \cite{WallaceGutzler1981, BarnstonLivezey1987, Cassou2008, GhilRobertson2002, FranzkeFeldstein2005}, which can have profound impact in improving predictability on
 intra-seasonal and longer, such as subseasonal to seasonal timescales \cite{ColucciBaumhefner1992}.

 Compared to the tropics, extratropical dynamics involves a great variety of wave-wave and wave-mean flow
 interactions, highlighting, hence, more involved nonlinearity, in particular, preferred intra-seasonal
 large scale structures of nonlinear flow regimes. The persistence timescale of these patterns is
 normally much longer than synoptic baroclinic timescales but smaller than typical intra-seasonal 
 variability timescales of radiative fluxes and bottom boundary conditions, such as sea surface 
 temperature anomalies \cite{BarnstonLivezey1987,Dole1986,Pandolfo1993}. 

 The extratropical persistent and quasi-stationary states are associated with teleconnection 
 patterns and states/positions of the jet stream \cite{Woollings+2010a,Woollings+2010b}. An extensive number of studies
has analysed and identified these nonlinear flow regimes ranging
 from various cluster analysis methods, bump hunting of the probability density function (pdf)
 to hidden Markov models and self-organizing maps. The number of these structures, however, is a 
 matter of debate between researchers, and depends on the location and extension of the geographical
 region and season. An extensive discussion with
 more details can be found in the review \onlinecite{Hannachi+2017}, and references therein.

 One of the challenging issues in the identification of the above extratropical nonlinear structures
 is the choice of the low-dimensional space, which allows appropriate reduction of weather-related 
 noise and efficient separation of these states. The space spanned by the leading empirical 
 orthogonal functions (EOF) is conventionally used as reduced space. EOFs have, however, a number
 of weaknesses putting limits on what can be achieved \cite{HannachiIqbal2019a, HannachiIqbal2019b}.
 To overcome those weaknesses and, in consistency with the nonlinear nature of the dynamics, these 
 authors applied kernel EOFs as a low-order state space to identify the nonlinear flow 
 structures. In particular, those structures are interpreted as  quasi-stationary states based on 
 the flow tendency within the same space. This flow tendency can only be applied with very long
 time series, as is possible with the quasi-geostrophic model of potential vorticity on the sphere
 but is not appropriate for data from reanalysis.

 Many problems in the real world, such as physical, computer, and social sciences, can be formulated
 and solved using the concept of networks or graph theory. A network is a collection of
 objects or nodes that are connected by edges. These connections can be defined
 based on a chosen metric in the system state space \cite{zou2019}. An example of such graphs can be found in the
 Isomap method \cite{Tenenbaum2000} and an application to the Asian monsoon can be found in 
 \cite{HannachiTurner2013}. Many networks allow a natural splitting of the system into 
 groups or communities/modules \cite{Newman2006}. To complement the analysis of the low
 frequency variability system within the low-dimensional kernel EOF space, we adopt and apply here
 for the first time the concept of network modularity to study the nonlinear dynamical feature of 
 the mid-latitude atmosphere.

 In this manuscript we revisit  and extend the analysis of Hannachi and Iqbal \cite{HannachiIqbal2019a} by using
 kernel EOFs combined with a recurrence network partitioning method \cite{Newman2006,marwan2009b}. The recurrence network analysis allows an elegant and easy partitioning of the 
 state space into communities in a natural way, providing an efficient way to identify the nonlinear
flow structure within the low-dimensional kernel EOF space.
Moreover, this makes it possible to use the tools of recurrence quantification analysis (RQA) \cite{MARWAN2007} to study important dynamical features of the detected structures.

The manuscript is organised as follows. 
 Sect.~\ref{sect_meth} describes the methodology,  the data and calculation procedure are given in Sect.~3,
 Sect.~4 presents the results, and a summary and conclusion are provided in the last section.
\section{Methods}\label{sect_meth}
\subsection{Kernel principal component analysis}

Detecting the regimes of a given (dynamical) system's variability can be formulated as the problem 
of separating structures of related states in the phase space. Typically, when studying the atmospheric
dynamics, we have at our disposal multivariate time series of some physical variables, such as 
temperature, pressure, geopotential height, etc., on a spatial grid, ${\bf x}_t, t=1, \ldots N$, 
and hence, the structures of 
interest are embedded in a high-dimensional  space. The nonlinearity of the system 
makes traditional linear data decomposition methods, such as empirical orthogonal function (EOF) 
analysis, inefficient and sometimes inadequate in disentangling 
these structures. The reason is that they are not necessarily oriented along linear directions, but are possibly
lying on complex manifolds, and may be 
embedded in a very large number of principal components (PCs). A suitable approach to overcome this 
complexity is to construct a nonlinear embedding of the state space, through a high-dimensional 
 multivariate mapping 
 $\mathbb\varphi(\cdot)$, from the original state space into a new {\it feature} space, so that 
the structures 
would be captured and could be well-separated by a few PCs in the new feature space. 
 In this setting, the problem is 
reduced to linear PCA applied to the transformed time series $\mathbb\varphi(\mathbf x_t)$ of the 
 original  multivariate $d$-dimensional time series 
$\mathbf x_t$ with $t=1,\dots N$. A straightforward way to solve this problem -- via explicit assignment
 of the functions $\mathbf\varphi$ -- is in most cases impractical, because it is difficult to guess 
both the functional form and its dimension,  which are optimal for detecting the regimes. 
The {\it kernel trick} can solve this problem in an elegant way. The 
kernel function $K(\cdot,\cdot)$ can be defined as a scalar product in such a way that 
$K( \mathbf x, \mathbf y) = \mathbb\varphi( \mathbf x)^T \mathbb\varphi( \mathbf y)$, and can be 
chosen from a large family such as polynomials or Gaussian functions. The choice of $K(\cdot,\cdot)$ 
 then defines implicitly the mapping $\mathbb\varphi$, i.e., without an explicit expression of it, 
 which is generally very high (and may be even infinite) dimensional.
 Then, all we need to know for calculating PCs in the new feature space is the matrix of inner 
products 
$K_{ij}=\mathbb\varphi(\mathbf x_i)^T\mathbb\varphi(\mathbf x_j):=\sum\limits_l\varphi_l(\mathbf x_i)\varphi_l(\mathbf x_j)$. 
Hence specifying the mapping $\mathbb\varphi(\cdot)$ is not needed to perform PCA; we can just introduce the kernel function $K(\cdot, \cdot)$ that defines the dot products $K_{ij}=K(\mathbf x_i,\mathbf x_j)$, $i,j=1, \ldots N$.
Such an implicit kernel-based nonlinear transformation of the original space is the core idea of the 
kernel PCA (KPCA) approach \cite{Scholkoph1998, Boser1992, Hannachi21}, which was shown to be effective, e.g., in 
identifying the LFV regimes in the extratropical  atmosphere \cite{HannachiIqbal2019b}. 
 The kernel function can be 
selected based on some general assumptions reflecting the similarity of states within the state space.
 According to the spectral decomposition theorem, 
the kernel function may be decomposed into an infinite series as 
$K(\mathbf x,\mathbf y)=\sum\limits_l\lambda_lf_l(\mathbf x)f_l(\mathbf y)$, 
where $f_l(\cdot)$, $l=1, 2, \ldots$, 
are the eigenfunctions of the integral operator with kernel $K(\cdot,\cdot)$. Accordingly, 
the mapping $\mathbb\varphi_l(\cdot)$, $l=1, 2,\ldots$, are then given by
$\mathbb\varphi_l(\mathbf x):=\sqrt{\lambda_l}f_l(\mathbf x)$, 
as mentioned 
in Ref. \onlinecite{HannachiIqbal2019b}. Thus, the approach makes it possible to consider infinite-dimensional 
embedding to achieve optimal  separation of distinct states.     
Technically, in  kernel PCA, the $N \times N$ matrix $\mathbf K$ is decomposed as follows
\begin{equation}
\mathbf K=\overline{\mathbf K}+\mathbf K_c=\overline{\mathbf K}+\sum\limits_{i=1}^{N-1} \mathbf u_i\cdot \mathbf u_i^T.  
\end{equation}
where the mutually orthogonal vectors $\mathbf u_i$, $i=1, \ldots N-1$, are 
 the kernel principal components (KPCs), and 
$\overline{\mathbf K}$ is 
%
\begin{equation}
\overline{\mathbf K}=\mathbf K-\mathbf K_c=\\
\frac{1}{N}(\mathbf{1\cdot K+K\cdot 1})-\frac{1}{N^2}\mathbf{1\cdot K\cdot 1},
\end{equation}
representing the deviation of $\mathbf K$ from the centered matrix  
 $\mathbf K_c=\mathbf{C\cdot K\cdot C}$, with  $\mathbf C$ being the $N \times N$ centering 
 matrix
 $\mathbf{C}=\mathbf{I}-\frac{1}{N}\mathbf{1}$,
and $\mathbf I$ and $\mathbf 1$ are respectively the identity 
matrix and the matrix of the same size filled with ones.  
The centering of the kernel matrix excludes the temporal mean of the features 
 (or states) in the feature space from the decomposition,
 which could result in a distortion of the 
decomposition as the leading KPC gets attracted towards the main diagonal.
This allows us to treat $\mathbf K_c$ as the matrix of covariances between the states at different 
times (temporal covariances) yielding, in particular, zero-mean of the KPCs.  
Thus, the KPC vectors $\mathbf u_i$, $i=1, \ldots N-1$,
  can be obtained from the eigendecomposition of the 
centered array $\mathbf K_c$: 
%
\begin{eqnarray}
\nonumber
\mathbf K_c=\mathbf{V\cdot D\cdot V}^T,\\ 
\mathbf u_i=D_{ii}^{1/2}\mathbf v_i, i=1, \ldots N-1
\end{eqnarray}
%
where $\mathbf v_i$ is the $i^\text{th}$ eigenvector, forming the matrix $\mathbf V$,   
and  $D_{ii}$ is the corresponding eigenvalue -- the variance of the $i^\text{th}$ KPC.  

Following the work \onlinecite{HannachiIqbal2019b}, here we use kernels that are Gaussian function of a 
distance $d(\cdot,\cdot)$ between state vectors at different times, i.e.,
\begin{equation}
\label{kernel}
K_{ij} = K(\mathbf x_i,\mathbf x_j)=\exp\left({-d^2(\mathbf x_i,\mathbf x_j)/2\sigma^2}\right).    
\end{equation}
Such a distance-based Gaussian kernel accounts for local similarity between 
 the states,
 which is a useful property for capturing nonlinear manifolds in the phase 
space.  The only generalization allowed in Eq.(\ref{kernel}), compared to the kernels used in Ref.
\onlinecite{HannachiIqbal2019b}, is the use of an arbitrary metric (not necessarily Euclidean) 
 determined by the specific problem.
 However, using some metric $d(\cdot,\cdot)$, we should ensure that the 
kernel function Eq.(\ref{kernel}), and, hence, 
 the matrix $\mathbf K$, are positive semi definite, since they 
are designed to define an inner product. This requirement is fulfilled with those metrics 
for which the metric space can be embedded in the Euclidean space \cite{Schoenberg1938}. 
 In case of 
other metrics, when negative eigenvalues of the kernel matrix are possible\footnote{
 this situation is out of scope in this article},
 we may consider using an \textit{approximation} of $\mathbf K$ by a 
positive semi definite matrix instead.

By applying KPCA to multidimensional time series we can expect clustering of the states 
in a space with low- to moderate number of KPCs, so that each cluster can be associated with 
 certain circulation regime of 
variability. The problem here is that neither the number of clusters, nor the dimension of the 
subspace in which the clusters are embedded, are \textit{a priori} known. This means that these 
parameters should be optimized for obtaining statistically justified clustering.  However, reliable 
optimization of the clustering procedure is difficult in real climate applications, due to 
insufficient statistics from the limited observed time series. Moreover, the clusters can 
have substantially non-Gaussian shapes, thus making such robust methods as, e.g., Gaussian mixture 
models \cite{Reynolds2009} or kernel density estimate \cite{Silverman86}, inefficient. Below we 
describe a method providing the detection of significant regimes that avoids such difficulties.            

\subsection{Recurrence network partitioning}

 Conventional recurrence networks are based on neighborhood thresholding using a Euclidean metric
 between pairs of states \cite{zou2019}. Given a set of multivariate states $\mathbf x_t$, $t=1, \ldots N$,
 and a recurrence threshold $\varepsilon$, the recurrence matrix $\mathbf R = (R_{ij})$ is defined by
 $R_{ij} = 1_{ \| \mathbf x_i - \mathbf x_j \| < \varepsilon}$, that is 
 1 if $\| \mathbf x_i - \mathbf x_j \| < \varepsilon$, and zero otherwise \cite{MARWAN2007}.
 In this regard the kernel matrix $\mathbf K$ can be used to produce a recurrence matrix through binarization
 using the metric $d(\cdot,\cdot)$ and threshold $\gamma$ as:  
%
\begin{equation}\label{rp}
R_{ij}(\gamma)=\begin{cases}
1,~ K(\mathbf x_i,\mathbf x_j)>\gamma \\
0,~ \text{otherwise}.
\end{cases}
\end{equation}
This matrix can be visualized as a recurrence plot (RP), by plotting $R_{ij}=1$ as a black pixel (and blank elsewhere). 
A line is then defined as a sequence of successive black pixels. 
A recurrence network is a graph using $\mathbf R$ as the adjacency matrix. The nodes of 
the graph correspond to the observed states $\mathbf x_t$, and if two states 
$\mathbf x_i$ and $\mathbf x_j$ are neighbors with respect to the metric $d(\cdot,\cdot)$, then the corresponding  nodes are connected by an edge, i.e., $R_{ij}=1$. 
The number $k_i = \sum_{j=1}^N R_{ij}$ is the degree of node $i$ and
represents the number of nodes connected to it (i.e., the number of recurrences
of the state at time $i$).
With this conceptual framework, the problem of 
regime detection can be formulated as recognizing communities of nodes, such that there are 
significantly more connections within communities than between them -- a situation akin to $k$-means
 clustering concerning between- and within-variances. Actually, each community 
joins the related states of the system based on the similarity measure
 Eq.(\ref{kernel}). Therefore, 
dividing the network into communities allows matching each state to a certain type of behavior.          

To detect the communities we use an approach suggested by Newman\cite{Newman2004}, in which the best 
division of the network maximizes a special cost-function called modularity. For a given 
division, the modularity measures the difference between the fraction of edges falling within the 
communities and the same fraction expected from a network with randomly distributed connections, 
regardless of the division. This random network is assumed to have the same number and degrees of 
nodes as in the analyzed network. 
Since elements of the matrix $R$ can only take 0 or 1, the expected value of $R_{ij}$ in a network with random connections equals to the probability to find an edge between nodes $i$ and $j$. This probability is estimated as $k_ik_j/2m$, where
$m=\frac{1}{2}\sum\limits_{i=1}^{N}k_i$, and represents 
the total number of edges in the network.
The modularity is then expressed as
\begin{equation}
\label{modularity}
Q=\frac{1}{2m}\sum\limits_{i,j}\left(R_{ij}-\frac{k_ik_j}{2m}\right)g_{ij}=\frac{1}{2m}\sum\limits_{i,j}\overline{R}_{ij}g_{ij},    
\end{equation}
with $g_{ij} = 1$ if nodes $i$ and $j$ belong to the same community and 0 otherwise. Here the matrix $\overline{R}_{ij}$ represents the deviation of $R$ from the expected adjacency matrix of a random network.

An elegant way to find the communities maximizing Eq.~(\ref{modularity}) was proposed in Ref.
\onlinecite{Newman2006}. The approach is based on iteratively splitting each community 
into two communities so that each split must provide the maximal positive increment of the whole 
network modularity, until indivisible communities are obtained. Let us consider a particular 
community $\mathbf H$ -- a subset of nodes of our network -- which we wish to split. 
If we are, for example, at the starting point of the algorithm, then 
 $\mathbf H$ is the whole set of nodes 
 indexed by $i=1,\dots, N$. Splitting $\mathbf H$ into two groups can be represented by a
 vector $\mathbf s$ (classifier or indicator), 
 whose elements $s_i=-1$ for the first group and $s_i=1$ for the second. 
 Note that the dimension of $\mathbf s$ is the size $i_H$ of $\mathbf H$.
 It can be shown that the increment $\Delta Q$ of the whole network modularity, after this split, 
 takes the form:
\begin{eqnarray}
\label{incr_mod}
\Delta Q=\frac{1}{4m}\mathbf s^T\mathbf B\mathbf s=
\frac{1}{4m}\mathbf s^T\mathbf W\Theta\mathbf W^T\mathbf s,\\ \nonumber
\mathbf B = (B_{ij}), \mbox{ with } 
  B_{ij}=\overline{R}^{(H)}_{ij}-\delta_{ij}\sum\limits_{k\in \mathbf H}\overline{R}^{(H)}_{ik},
\end{eqnarray}
%
where $\overline{\mathbf R}^{(H)}$ is the submatrix of $\overline{\mathbf R}=(\overline{R}_{ij})$ 
obtained by selecting
the elements of $\overline{\mathbf R}$ with the indices $i,j\in\mathbf H$, $\delta_{ij}$ is the 
Kronecker delta, and $\mathbf W\Theta\mathbf W^T$ is the eigendecomposition of the matrix $\mathbf B$,
with $ \Theta = \text{diag} ( \theta_0, \ldots, \theta_{i_{\mathbf H}-1})$. 
 Accordingly, the problem of splitting $\mathbf H$ boils down to finding the classifier (or indicator) 
 vector $\mathbf s$ consisting of numbers $1$ 
 and $-1$ that maximizes the quadratic form, Eq. (\ref{incr_mod}), which is equivalent to maximizing the 
 dot product of $\mathbf s$ with the eigenvector $\mathbf w_0$ of the matrix $\mathbf B$ corresponding 
 to its largest eigenvalue $\theta_0$. 
 The exact solution is the vector $\mathbf s$ with  
 components $s_i$ having the same sign as the corresponding components  $w_{0i}$ of the 
 leading eigenvector $\mathbf w_0$ of $\mathbf B$. 
 If the matrix $\mathbf B$ has positive eigenvalues, two new communities 
 defined by the vector $\mathbf s$ will emerge, provided that $\Delta Q>0$. Otherwise, if there are no
positive eigenvalues in $\mathbf B$, its largest eigenvalue is always zero, since one of the 
properties of this matrix is the zero sum over each row or column. In this case, all components of 
the leading eigenvector have the same value, which means that all $s_i$ are also the same, 
i.e. $\mathbf H$ is no longer divisible. The described splitting process of the network communities 
can continue until the $\Delta Q>0$ condition is violated for each current community\footnote{In
  practice, 
however, we may consider using a threshold slightly above zero to avoid ``too thin'' separation, if, 
for example, further splitting gives much less modularity increment as compared with the previous 
splits}.                

Besides the indicator or classifier vector $\mathbf s$, useful information about the structure 
 of the resulting communities is also provided by 
 the leading eigenvector $\mathbf w_0$. As it can be seen from Eq.(\ref{incr_mod}), the absolute 
value of $w_{0i}$ measures the contribution of the $i^\text{th}$ element to the modularity of the network, i.e., 
 the gain in modularity from the inclusion of the corresponding node in the community. 
 The number $|w_{0i}|$ is referred to as the {\it centrality} of the 
 corresponding node in the resulting community, and therefore 
 for each community we get a vector of centralities. In terms of interpretation in connection to
  the atmospheric circulation
 regime detection, large centrality of some node $k$ of the network indicates that the 
corresponding spatial pattern $\mathbf x_k$ is similar to a large number of other patterns belonging 
to the same community, and represents therefore a typical pattern for the regime associated with this 
community.   

The main advantage of Newman's method described above is that it provides an efficient and elegant
way to cluster the network into communities without any prior information on their size and number
 using simple 
matrix calculations. 
 This is particularly convenient in climate data analysis where the set of regimes is not known 
in advance and the 
problem of the data-driven identification comes to the forefront.
 The suggested method  of decomposing  the analyzed time series allows us to label the states 
of the system in the space of the leading  kernel PCs in accordance with the detected communities.
As a result, looking at the labelled/marked states in the KPC space, 
we can easily select 
{ a number of leading KPCs that provides clear separation of the communities}
(see the 
results below { and a simple example in Appendix}). Further, the selected KPCs can play the role of dynamical variables that describe 
the sporadic switching of the system trajectory between the dynamical circulation regimes.    

\subsection{Studying dynamical properties of the regimes}\label{sec:methods:recurrence}
The adjacency (recurrence) matrices of the partitioned recurrence networks can be analysed using RQA \cite{MARWAN2007}.
Each regime-specific recurrence plot (RP) 
can be regarded as a particular subset of the full RP constrained by the community labels. In order to 
compare the dynamical properties of the different regimes, RQA is carried out separately for each regime-specific RP.


A simple quantifier for the overall intrinsic \textit{similarity} of an atmospheric regime can be defined by the ratio of total 
recurrences in the regime-specific RP relative to the (squared) time that is spent in the respective regime, i.e., recurrence rate 
RR, $RR =\sum_{i,j} R_{ij}/N^2$.
In traditional RQA, the statistics of diagonal and vertical line structures are studied to characterize the predictability and 
intermittency of a system. Diagonal lines in an RP reflect time periods during which two segments of the phase space trajectory 
evolve in parallel, indicating deterministic and well-predictable dynamics. By counting the number of diagonal lines that exceed 
a specific length for the regime-specific RPs, we study if the evolution of spatial patterns during different time periods is 
similar. \textit{Predictability} of atmospheric patterns in each regime is quantified by the determinism DET of the RP which is 
given as the fraction of diagonal lines that exceed a minimum line length of $l_{\mathrm{min}} = 5$ 
days to all diagonal lines, i.e. 
$ DET = \sum_{l\ge l_\text{min}} lP(l)/\sum l P(l)$, where $P(l)$ is the distribution of diagonal line lengths $l$.
Vertical lines in an RP identify periods in which the dynamics are `slowed down’. Consequently, we interpret them as 
corresponding to quasi-stationary patterns potentially associated with atmospheric blocking and possibly zonal flow. 
\textit{Persistence} of atmospheric patterns in each regime is quantified by laminarity LAM which is computed as the fraction of 
the total vertical lines with $l_{\mathrm{min}} = 5$ 
days to all vertical lines: $LAM= \sum_{v \ge v_\text{min}} vP(v)/\sum vP(v)$,
where $P(v)$ is the distribution of vertical line lengths $v$.
Recently, an approach using recurrence lacunarity (RL) was proposed to characterize features of an RP that are distributed among 
multiple time scales and are not neccessarily expressed in line structures \cite{braun2021detection}. RL generally reflects the 
heterogeneity of an RP. Thus, we interpret it as the 
\textit{diversity}
{ of the regime behavior}.  
While regular RL informs about the general heterogeneity of recurrences, its extension 
to diagonal/vertical line structures is straight-forward and is introduced here. For the computation of diagonal/vertical line RL 
(dRL/vRL), the number of diagonal/vertical lines exceeding $l_{\mathrm{min}} = 5$ 
days
in each box is counted and reflects how strongly predictability/persistence of atmospheric patterns varies throughout different 
time periods. The box width is fixed to one year, highlighting interannual variability. High values indicate that, e.g., high 
persistence during one time period only has limited implications for other time periods. Thus, we always show these three 
different RL-based measures of diversity.

Significance testing allows to test the dynamical properties of atmospheric regimes against different null hypotheses. We test 
for two different hypothesis: (i) we check whether the recurrence network partitioning yields regimes that are significantly 
different from random regimes with respect to above mentioned recurrence quantifiers and (ii) 
we test which regime yields significantly high values for a given recurrence quantifier.

The first test is done by random deletion of recurrences from the full RP. In particular, for the $i^\text{th}$ regime-specific 
RP with $n_i$ recurrences, we randomly delete $m = N - n_i$ recurrences from the full RP that contains $N>n_i$ recurrences while 
also reproducing the column-wise recurrence rate of the $i^\text{th}$ regime-specific RP. We generate 200 random samples for each 
RP and compute the $99$\%-quantile for each RQA measure from this ensemble as an upper confidence level to test for 
significance.

For the second test, we apply a bootstrapping procedure with $n_B=2,000$ runs \cite{marwan2013recurrence}: we first collect all diagonal lines(/vertical lines/counts) from the 
distributions obtained from each regime-specific RP separately. In a single bootstrap run, we draw $M$ times (with replacement) 
from the unification of these length(/count) distributions and compute the measure of interest, yielding a single value. $M$ is 
given by the number of lines/the total count in the $i^\text{th}$ regime-specific RP.
By repeating this procedure $n_B$ times, an empirical test distribution for this measure is obtained from which we compute the 
upper $99\%$-confidence level. This method is applied for all measures except the recurrence rate; in this case, we test against the 
hypothesis that recurrences are distributed among the regimes with respect to the time spent in each regime. Consequently, we 
obtain each regime-specific significance level by dividing the total number of recurrences (from all regimes) by the individual 
(squared) time spent in a regime.

\subsection{Method summary}
\begin{figure}[ht!]
\includegraphics[width=1 \linewidth]{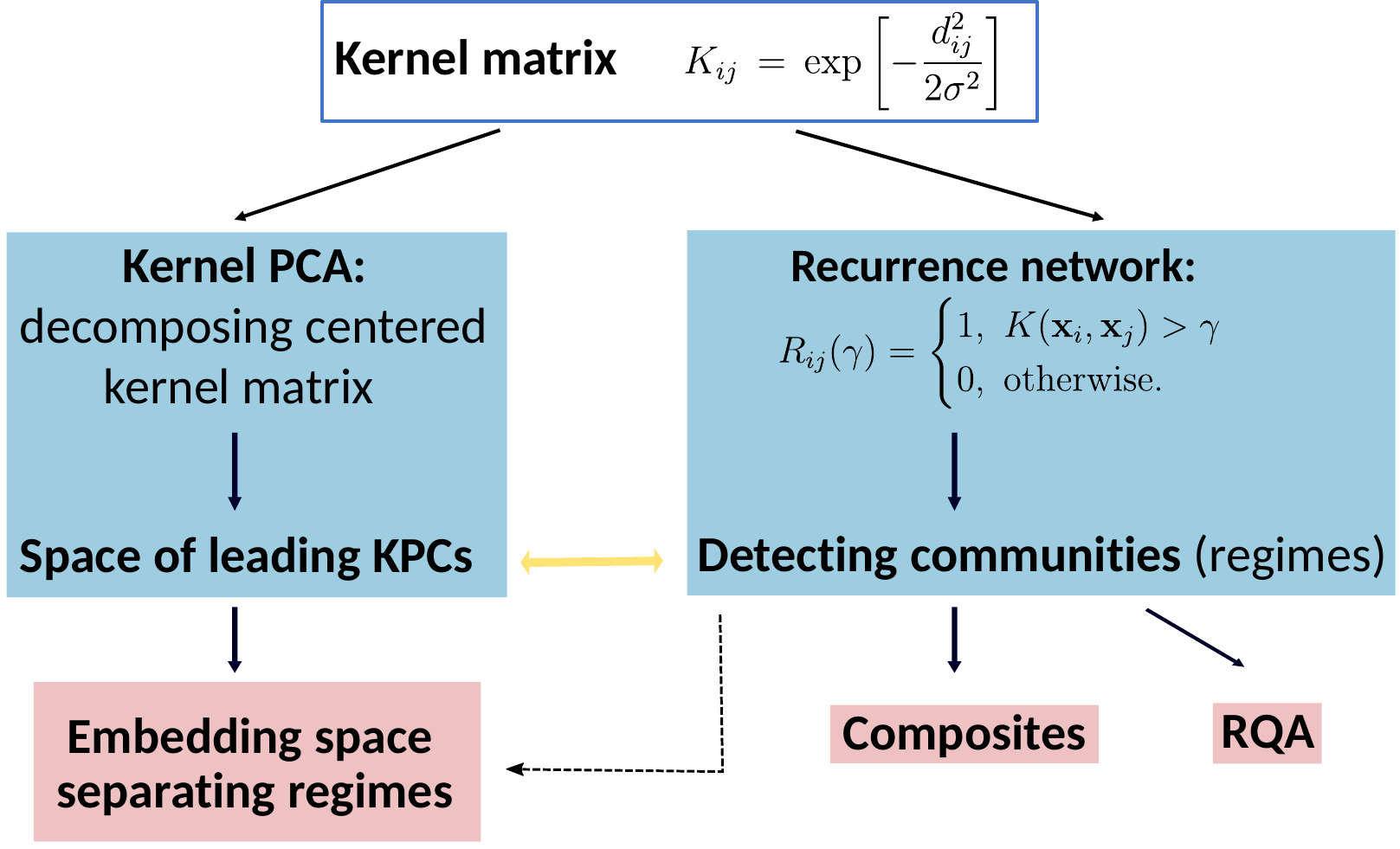}
\caption{Schematic representation of the proposed procedure.}\label{fig:flowchart}
\end{figure}

{ The whole procedure described in the above subsections A-C is represented schematically in Fig. \ref{fig:flowchart}.
The main goal of the KPCA is to obtain a few number of variables (leading KPCs)  which are suitable for differentiating the obtained regimes. The same kernel matrix as in KPCA is used for building the recurrence network which partitioning yields a set of regimes. As a result, we obtain a state space in which the regimes are well separated as well as representation of the regimes in physical space via composites, and can study properties of the regimes by the RQA methodology.} 

\section{Data and calculation setup}
\label{sec:data}
\subsection{QG3 model time series}
Quasi-geostrophic (QG) models of the atmosphere are popular polygons for testing algorithms 
concerning weather/climate dynamics.
Being simplified representations of atmospheric dynamics, QG models demonstrate a rich 
spectrum of variability at different time scales and are competitive to 
intermediate and full general circulation 
models  regarding complexity and  dynamical features/processes. 
Here we use time series generated by a three-level QG model (QG3) on the sphere
\cite{Marshall1992} with realistic 
orography and surface boundary condition. Based on the potential
vorticity equations at three (200, 500, and 800 hPa) pressure levels
\cite{Marshall1992,Vannitsem1997,Corti1997}, the model
is tuned to simulate winter atmospheric circulation over the extratropical hemisphere. The model 
exhibits highly nonlinear behavior with a chaotic attractor in its phase space with more than 
100 positive Lyapunov exponents \cite{Vannitsem1997}. The LFV behaviour of the model 
regimes was studied in a 
number of works, based on different kinds of clustering in a truncated phase space. In 
particular, the authors of the work \onlinecite{Kondrashov2003} identified four clusters in the space of three leading 
PCs calculated from a very long (54,000 days) time series of the mid-level stream function anomalies (SFA). 
These clusters are associated with the well-documented atmospheric modes or teleconnections,
 namely, Arctic oscillations (AO) and 
North Atlantic oscillation (NAO). Similar results were obtained in Ref. \onlinecite{Seleznev2019} based on 
reduced data-driven models, but using a much shorter sample of 5,000 days. 
Hannachi and Iqbal\cite{HannachiIqbal2019b} used KPCs as a space for a PDF-based cluster detection; however, only two 
clusters related to the AO were detected. 
Here we present results of our analysis applied to three 
10,000-day time series of the mid-level stream function anomalies, distributed over latitudes 
36\textdegree N to 90\textdegree N with approximately $5.5 \times 5.5$ degree resolution. 
These non-overlapping time series are randomly taken from a very long (300,000 days) QG3-model run. { Then the analysis is performed to each time series independently.}  

\subsection{Reanalysis data}
\label{sec:data_rean}

To study the circulation regimes of the real atmosphere, we use daily geopotential height (HGT) 
time series at the 500 hPa pressure level from the NCEP/NCAR reanalysis dataset \cite{Kalnay1996}. 
The time series is provided on
a $2.5 \times 2.5$ degree latitude-longitude resolution, north of 30\textdegree N covering the period 1980 to 2020.
The data are de-seasonalized by removing the daily seasonal mean
signal over the whole period smoothed in time with a Gaussian 
window with the standard deviation of 15 days.
Such smoothing suppresses the day-to-day noise in the resulting annual cycle, while re-taining the 
intra-annual seasonal structure. 
Only winter (December-January-February) values are taken from the obtained time 
series of daily HGT anomalies, yielding a sample of 3,579 days.  

\subsection{Distance metric, kernel and recurrence matrices}

\begin{figure}[ht!]
\includegraphics[width=0.7 \linewidth]{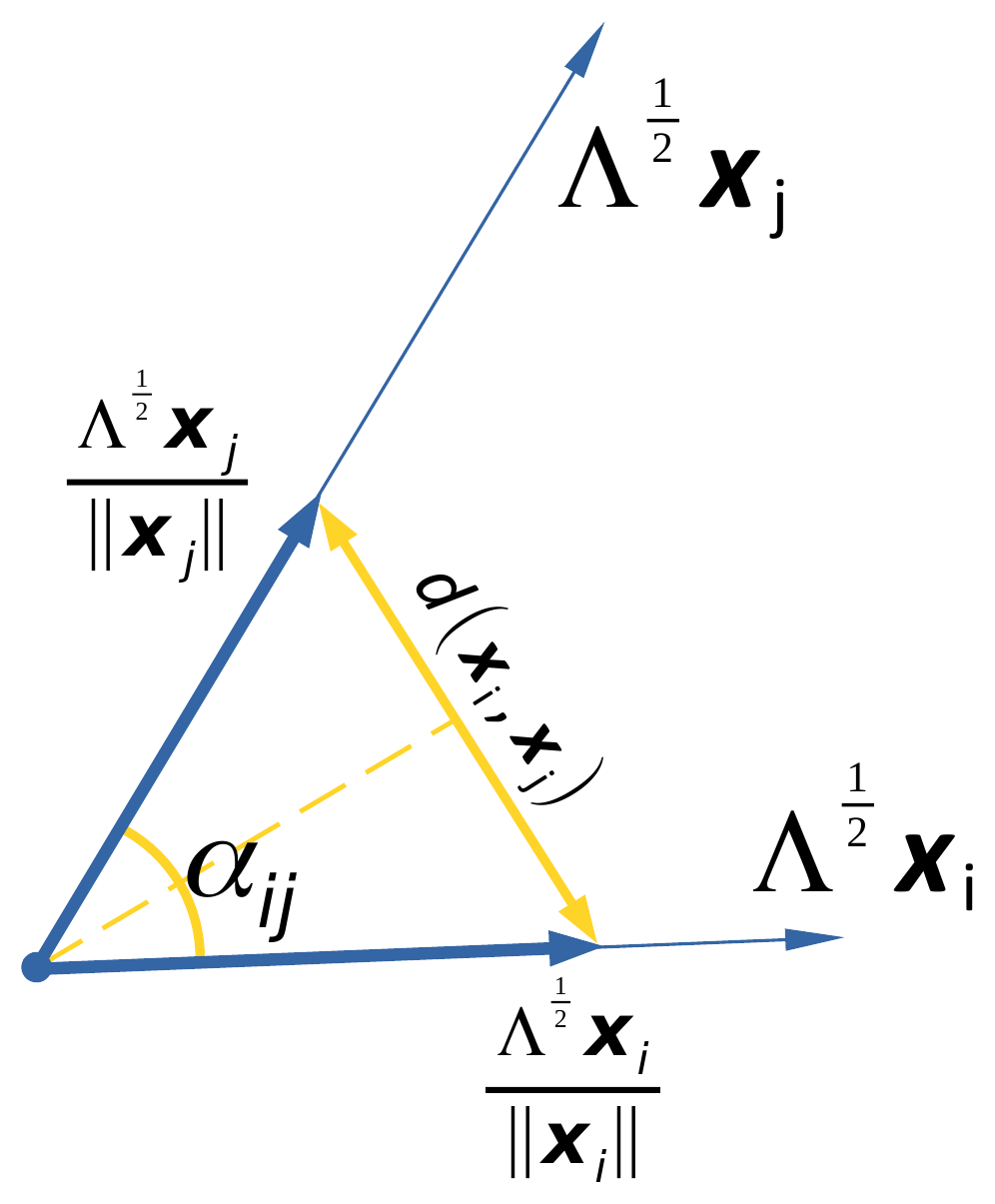}
\caption{Representation of the distance defined by Eq. \ref{distance}.}\label{fig:sketch}
\end{figure}

Since our goal is to separate sets of states, each of which joins spatial patterns 
yielding similar atmospheric conditions, we focus on the pattern's structure/shape rather than 
their amplitudes. Accordingly, we define a distance given by the sine of the angle $\alpha_{ij}$
between pairs of patterns: 
%
\begin{equation}\label{distance}
d(\mathbf x_i,\mathbf x_j)=\left\lVert\frac{\mathbf x_i}{\lVert\mathbf x_i\rVert}-\frac{\mathbf x_j}{\lVert\mathbf x_j\rVert}\right\rVert = 2 \left| \sin \frac{ \alpha_{ij}}{2}\right| ,    
\end{equation}
where $\lVert\mathbf x\rVert=(\mathbf x^T\Lambda\mathbf x)^\frac{1}{2}$ is the metric using a $d \times d$ diagonal 
weighting matrix $\Lambda$ reflecting the non-uniformity of the spatial grid (see below). 
Clearly, this metric inherits all properties of the Euclidean metric, as it is nothing more than the
Euclidean distance between the weighted vectors $\Lambda^\frac{1}{2}\mathbf x$ normalized to unit 
norm (see Fig. \ref{fig:sketch}). This always yields a positive semi-definite kernel matrix Eq.~(\ref{kernel}).

A very important parameter that determines the recurrence network structure and strongly impacts 
the splitting of the network into communities is the threshold $\gamma$ in the definition of the 
recurrence matrix, Eq.~(\ref{rp}) \cite{marwan2011}. If it is too large (remember that here the thresholding is opposite to the classical recurrence
definition),
the network degenerates into many communities 
yielding high modularity. 
Such a network may eventually not help to reveal any connections between patterns except those close in time.
\begin{figure}[ht!]
\includegraphics[width=\linewidth]{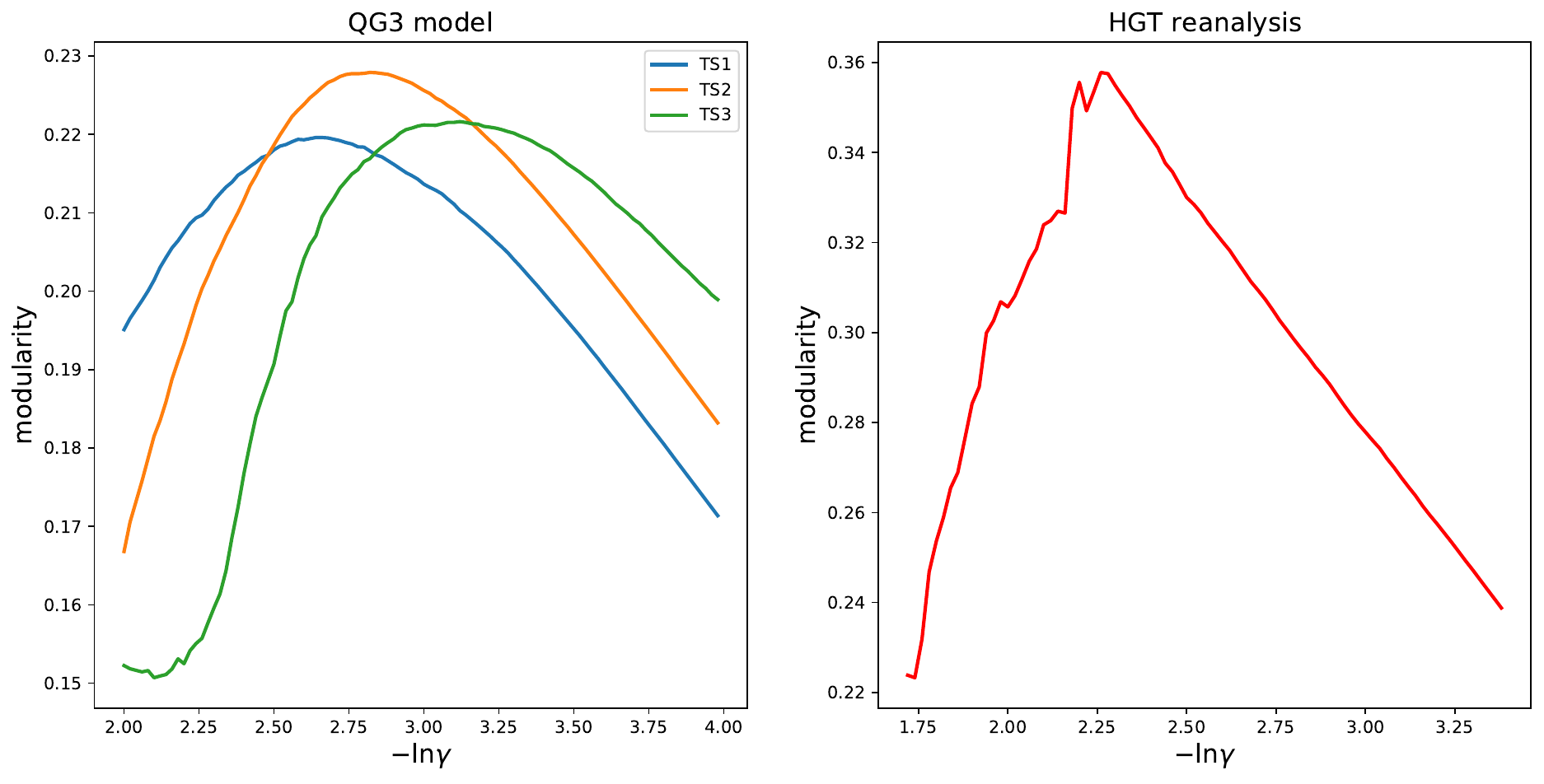}
\caption{Modularity increment after first division of the network into two communities vs. the threshold $\gamma$. Curves obtained from the three time series of the QG3 model and the HGT reanalysis time series are plotted in the left and right panels, respectively.
}\label{fig:mod_curve}
\end{figure}
If, on the contrary, $\gamma$ is too small, we end up with a 
poorly divisible network, in which each node is connected to most of the remaining nodes. Hence, this calls for the need to optimize the threshold parameter. 
Here we select $\gamma$ to provide the best division of the network 
into two basic communities at the initial splitting of the whole network. Under this requirement, 
the resulting value of $\gamma$ maximizes the modularity increment $\Delta Q(\gamma)$ at the first 
iteration of the algorithm. Such a choice is justified in the case of mid-latitude atmospheric 
dynamics, since two opposite types of circulation associated with the strength of the polar vortex 
are known to be the dominant modes in this region \cite{Thompson1987}, and the network should 
distinguish them well at the most basic level. It is obtained (see Fig.~\ref{fig:mod_curve})
that there are 
pronounced maxima of this dependence for both datasets considered here.  By adjusting $\gamma$ 
we do not need to care too much about the precise value 
 of $\sigma$ in the kernel function, Eq.~(\ref{kernel}). 
We find that setting $\sigma=2\min\limits_{i,j}d(\mathbf x_i,\mathbf x_j)$ in all examples below 
provides quite robust results, which is close to assumptions from the work \onlinecite{HannachiIqbal2019b}. 

\begin{figure}[ht!]
\includegraphics[width=\linewidth]{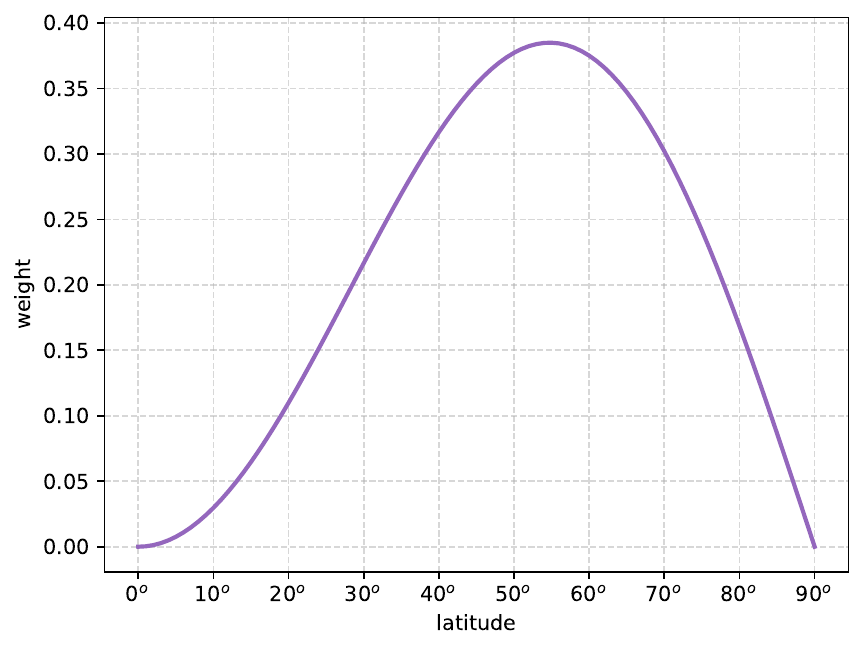}
\caption{Dependence of the weight given by Eq.~(\ref{weight}) on latitude.}\label{fig:weight}
\end{figure}

\subsection{Data weighting}

The data considered here is defined on a grid that is uniform in polar coordinates, so its 
cells cover unequal areas.
In line with the traditional approach, we had to distribute the weights of the grid nodes in the 
distances $d(\mathbf x_i,\mathbf x_j)$, Eq.(\ref{distance}), according to the area fractions around  each grid node. This corresponds to 
weights $\Lambda_{mm}\propto\cos{\theta_m}$, where $\theta_m$ is the latitude (in radians) of the $m^\text{th}$ component $x_m$ (or $m^\text{th}$ grid point) $m=1,\ldots d$, of the state vector ${\bf x}$. A main problem with such weighting in our case is 
 related to the resulting larger magnitude of anomalies at the 
southern bound of the considered latitude band. This moves the focus of analysis from the 
mid-latitudes -- the region of interest -- to the subtropics.
 No weighting, on the other hand, yields dramatic increase 
 of polar latitudes contribution, thus shifting  focus on the 
Polar vortex region rather than the mid-latitude circulation. 
The problem of proper weighting was stated, e.g., in the works \onlinecite{HainesHannachi1995,Hannachi1997}.
Here, 
we take into account the latitudinal dependence of the characteristic spatial 
scale of atmospheric anomalies responsible for planetary-scale circulation regimes. In the 
extratropics, this scale is proportional to the Rossby deformation radius \cite{Pedlosky1987}, 
which depends on the latitude as $\frac{1}{\sin{\theta}}$. Therefore, normalization of the grid cell 
areas  through scaling by the characteristic areas of anomalies ($\propto\frac{1}{\sin^2\theta}$) 
allows us to better capture the peculiarities and typical features of the atmospheric circulation 
 using a given spatial grid. 
As a result, we use weights in the distance measure that focus on the midlatitudes 
expressed as 
\begin{equation}
\label{weight}
\Lambda_{mm}=\cos{\theta_m}\sin^2\theta_m,    
\end{equation}
 or, equivalently, multiply the signal at each grid point by 
 $\sin{\theta_m}\sqrt{\cos{\theta_m}}$. Such 
weighting distinguishes mid-latitudinal and subpolar nodes of the grid, reaching the maximum at 
approximately 55\textdegree N (Fig.~\ref{fig:weight}).


\begin{figure*}[ht!]
\includegraphics[width=\linewidth]{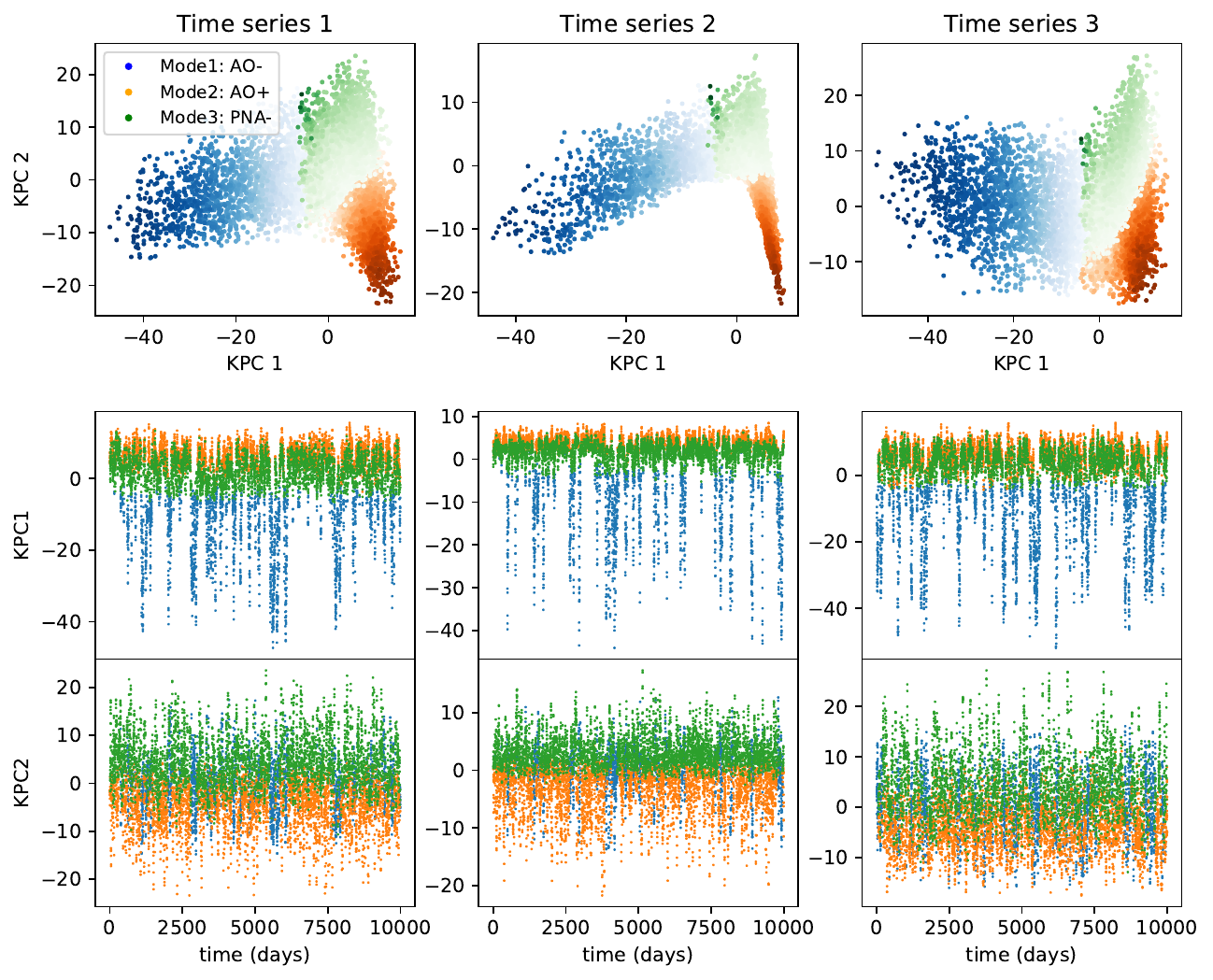}
\caption{Regimes of the QG3 model behavior in the space of the leading two KPCs. The three columns correspond to the three analyzed time series. In upper panels 
states of each regime (mode) are shown in the KPC1-KPC2 plane (see the text). 
States belonging to different regimes are marked by different colors; color saturation corresponds to the centrality of a state in its community. Time series of the KPC1 and KPC2 variables are plotted in the lower panels.}
\label{fig:QG3_kpcs}
\end{figure*}

\section{Results}
\label{sec:results}
\subsection{QG3 model time series}

We apply the methodology described 
above to each of the three SFA time series separately. For all time series the recurrence network division method gives the same 
number of communities, or regimes, equal to three. To illustrate clustering of the communities in the KPC-space, we plot 
elements of each community in the plane of the two leading KPCs (see 
Fig.~\ref{fig:QG3_kpcs}). The most typical (i.e., having large centralities in their communities) states from different communities are well-separated in this plane, and hence, these two variables 
can serve well as an embedding for the three identified regimes.

\begin{figure*}[ht!]
\includegraphics[width=\linewidth]{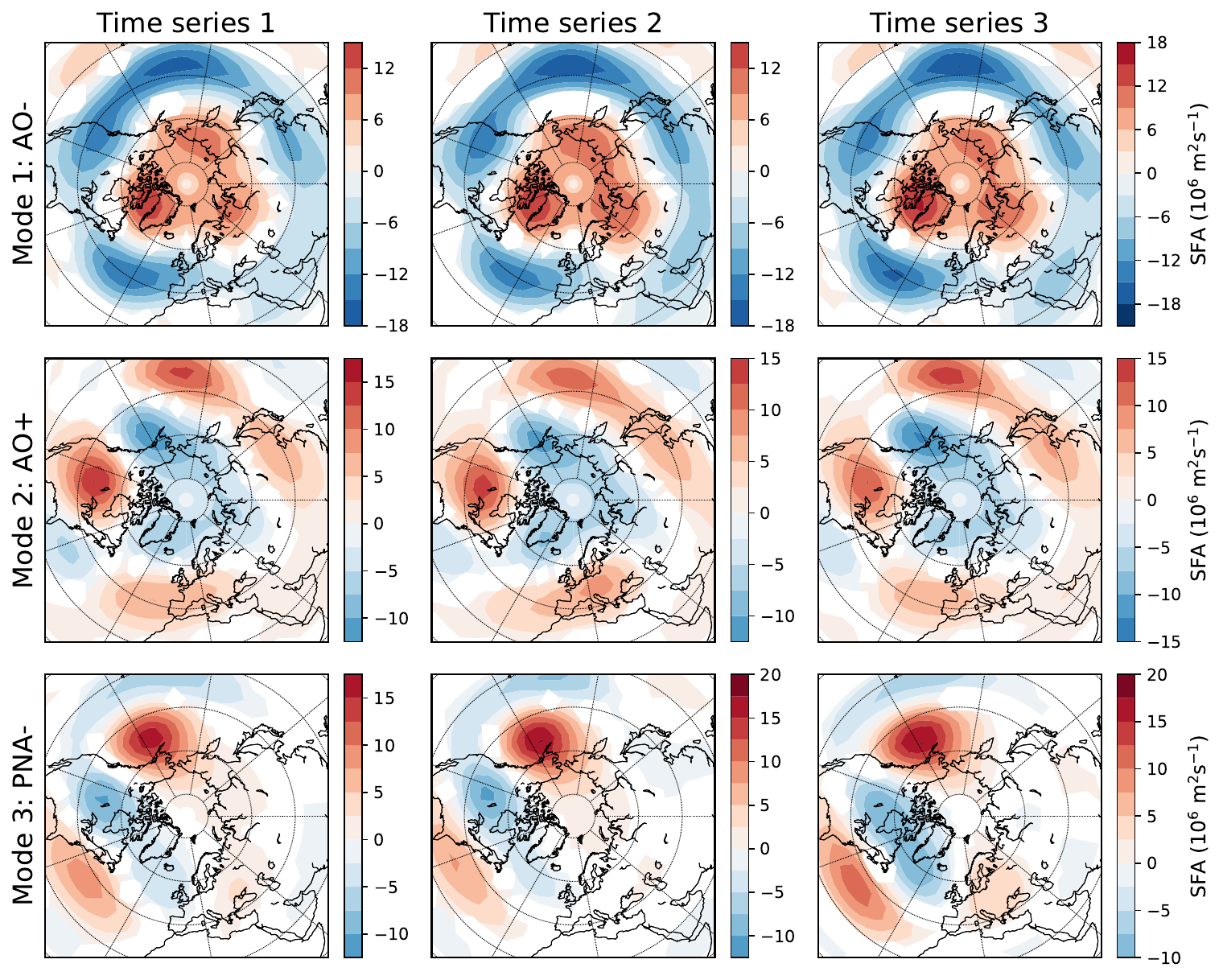}
\caption{Composite patterns of the QG3 model SFA corresponding to the obtained regimes. Columns correspond to different time 
series, rows to different regimes. The composites are calculated as SFA averaged over 20\% most typical states of each of the 
obtained regimes (see the text). { Only values that are significantly different from zero by Student's t-test with a critical value of 0.01 are shown.}}
\label{fig:QG3_comp}
\end{figure*}

Unlike in EOF decomposition or some of its nonlinear generalizations (see, e.g., Refs. \onlinecite{Mukhin2015,Mukhin2018}), there is no explicit 
mapping of the KPCs to the data space.   
A coarse geographical structure of SFA relating to a specified regime can be obtained via composite analysis, i.e., the spatial 
field of SFA averaged over the most central states of the regime (as shown in Fig.~\ref{fig:QG3_kpcs}, top). The resulting 
composites are clearly related to the three well-known atmospheric teleconnection patterns (Fig.~\ref{fig:QG3_comp}). The 
first two are the positive and negative phases of AO, connected with an anomalous pressure difference between the polar region 
and the mid-latitude belt, and, respectively, stronger or weaker (zonal) westerly flow. The third one resembles the negative 
Pacific North American (PNA) pattern characterized in winter seasons by dominating tripol structure with positive anomalies over 
the North Pacific and near southeastern United States and negative anomalies over central Canada. We note that we have not obtained 
a separate regime corresponding to NAO, but NAO-related anomalies are captured by mode 1, which encompasses the negative NAO phase, whereas modes 2 and 3 feature of the positive NAO phase. 

{ Studying the nature of, and the relation between the various teleconnection
patterns, including PNA, NAO and AO/NAM (Northern Hemisphere Annular Mode), has a long history (e.g. Refs. \onlinecite{WallaceGutzler1981,BarnstonLivezey1987}).
The nonlinearity and/or non-distinguishability between the NAO and AO
teleconnection has been discussed thouroughly in the literature (e.g. Refs. \onlinecite{Thompson1987,Wallace2000}).
The linear EOF analysis, and
even nonlinear cluster analysis cannot categorically distinguish between 
NAO and AO. For example, Feldestein and Franzke\cite{Feldstein2006} examined, based
on composite analysis the null hypothesis that the NAO and AO/NAM 
persistent events are not distinguishable. They found that the null hypothesis cannot be rejected even at 20\% significance level.
In another analysis, Dai and Tan\cite{Dai2017} examined the nature of AO through
SOM (Self Organizing Map) analysis. They found that AO, derived from the
250-hPa geopotential height anomalies, can be interpreted in terms of a continuum that can be approximated by five discrete AO-like pattern, which 
overlap with the discrete NAO-like pattern. These findings explain why 
separate NAO regimes are not identified in this analysis.
}

The negative AO regime is the most  distinguishable mode of the QG3 model dynamics (Tab.~\ref{tab:net_param}): 
its contribution to the network modularity is substantially greater than the contribution of the other two modes, although with a 
small frequency of occurrence. { For this reason, the most central states of this mode are well separated from other states, with excursions into the area of large negative values of the first leading KPC. } 
{ Consequently,} the transitions to this regime look like rare { irregular} outliers in the time series of the 
first KPC (see Fig.~\ref{fig:QG3_kpcs}, bottom). Indeed, this KPC1 can be considered as an index describing joint AO and 
NAO dynamics. At the same time, the second variable (KPC2) helps differentiate between the positive AO and negative PNA states 
(Fig.\ref{fig:QG3_kpcs} top), both of which contribute to the positive NAO.

\begin{figure*}[ht!]
\centering
\includegraphics[width=\textwidth]{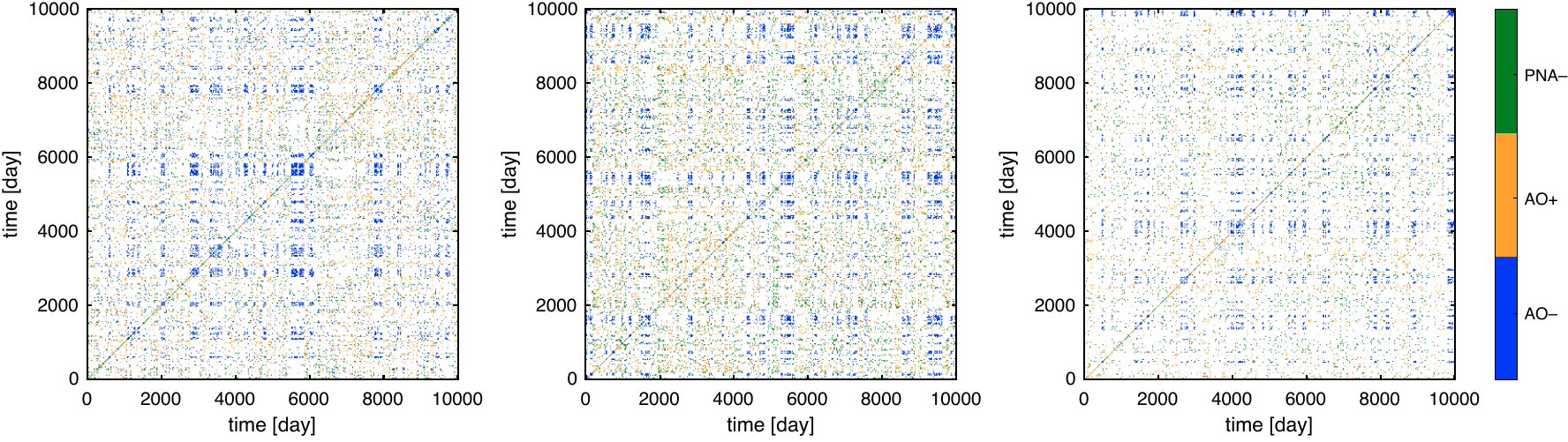}
\caption{Recurrence plots of atmospheric patterns obtained from the QG3 model data for the three different time series. Recurrences within a given regime are color coded accordingly. A specific rendering is used to make the RP appear less sparse.}\label{fig:rp_qg3}
\end{figure*}

\begin{figure}[ht!]
\includegraphics[width=\linewidth]{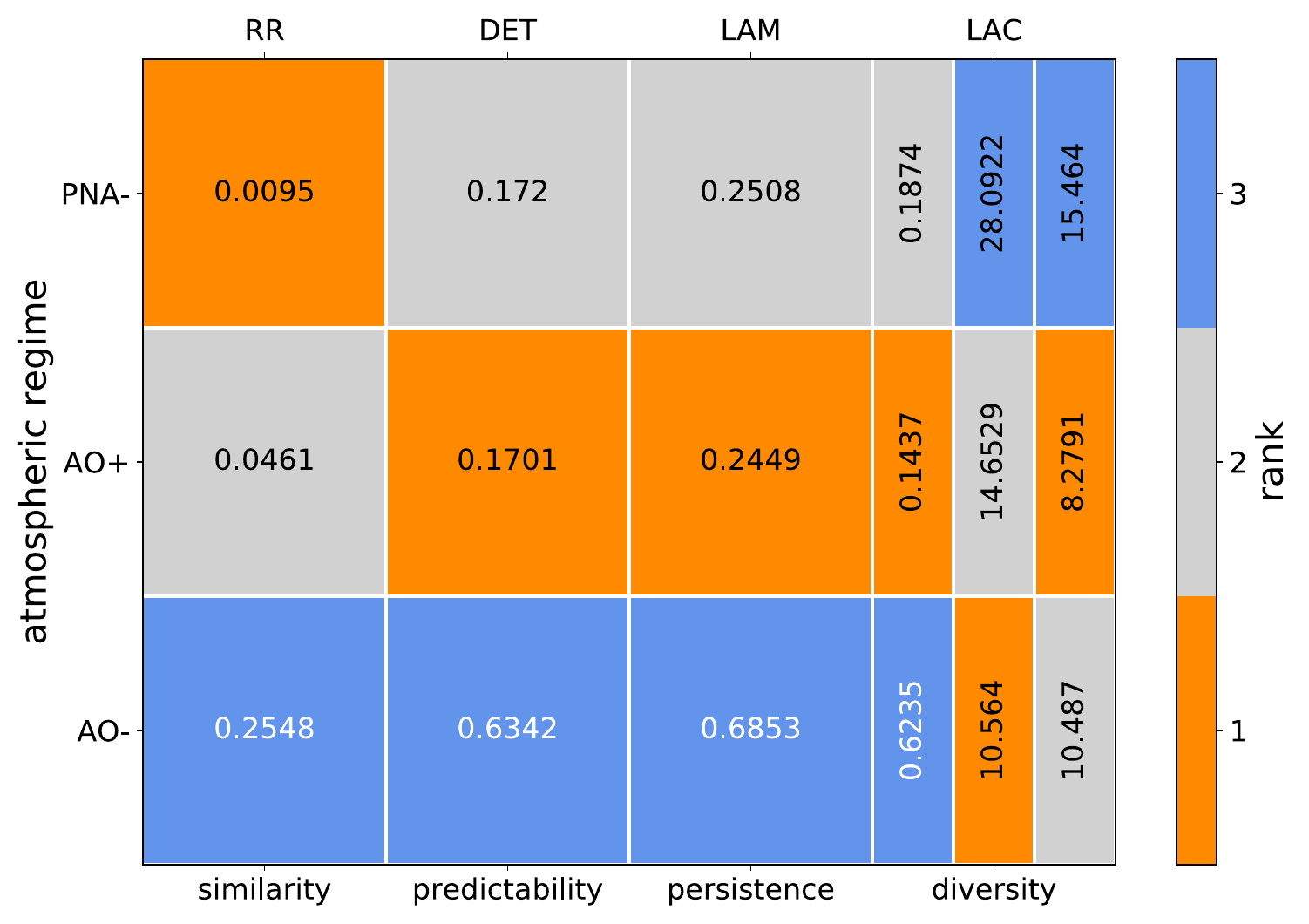}
\caption{Results of recurrence quantification analysis for the atmospheric regimes obtained in the QG3 model data set for one of 
the three time series. The four different RQA measures are labelled with their respective interpretation. Significant 
(insignificant) values are printed in white (black) numbers. Color coding illustrates the ordering of the RQA values (ascending 
column-ranking) for better comparability.}\label{fig:rqa_qg3}
\end{figure}

Let us now turn to the dynamical properties by considering the RPs separately for each regime (Fig.~\ref{fig:rp_qg3}, see. also Sect.~\ref{sec:methods:recurrence}), along with the RQA (Fig.~\ref{fig:rqa_qg3}).
The RQA reveals distinct dynamical properties of the identified regimes (Fig.~\ref{fig:rqa_qg3} and supplementary Fig. 1).
The negative AO regime 
stands out as the most similar, persistent and predictable regime. This suggests that it encompasses 
periods of atmospheric blocking as these are characterized by atmospheric patterns that persistently 
reside for relatively long time periods without significant spatial variations. During a blocking event, 
atmospheric conditions are consequently more similar and predictable. On the other hand, the negative AO 
regime is characterised by a high degree of 
diversity, 
reflecting the low predictability of blocking events on 
interannual and decadal time scales. These findings are fostered by the corresponding regime-specific RP 
(Fig.~\ref{fig:rp_qg3}); on longer time scales, the RP appears heterogeneous while block-structures reflect periods of atmospheric blocking.
The positive AO regime is associated with stronger westerly zonal flow which appears to result in 
transient, short-lived atmospheric patterns that do neither exhibit a significant degree of persistence
nor allow for reliable short-term predictions. Low-frequency variability of atmospheric patterns in this 
regime is relatively low as indicated by low values in the different RL measures.
Time periods, during which atmospheric conditions are characterized best by the negative PNA regime are identified with moderately predictable and persistent dynamics. Given the total time 
this regime is detected, the number of recurrences is relatively low, representing low similarity. Both 
diagonal and vertical line structures in the regime-specific RP exhibit strong heterogeneity, suggesting that the temporal variations in this regime run through both well-predictable, persistent and stochastic, 
volatile periods. This could be interpreted as a high degree of non-stationarity of the dynamical 
properties of this regime.
Finally, results between the three different time series show good general correspondence (supplementary Fig. 1),
supporting the given interpretation of regimes. The most significant deviation is found for the 
first time series compared to the second and the third with respect to the similarity of all three 
regimes.

Overall, the results (Figs.~\ref{fig:QG3_kpcs}, \ref{fig:QG3_comp}, Tab.~\ref{tab:net_param} and supplementary Fig. 1) confirm that the suggested methodology gives a fairly stable solution: the number of regimes, their spatial and temporal structures, the embedding spaces, as well as the modularity rates, show coherency between the three different, independently analyzed time series.      

\begin{table*}
\centering
\begin{tabular}{|l|l|l|l|}
\hline
~ & Community 1 (AO-) & Community 2 (AO+) & Community 3 (PNA-)\\
\hline
Modularity (TS1; TS2; TS3) & 0.219;~0.222;~0.228 & 0.169;~0.174;~0.155 & 
0.107;~0.118;~0.103 \\
\hline
Number of states (TS1; TS2; TS3) & 2235;~1861;~2484 & 3746;~3863;~4058 & 4019;~4275;~3458 \\
\hline
Modularity per state (TS1, TS2, TS3), ($\times 10^{-5}$) & 9.8;~11.9;~9.18 & 4.51;~4.5;~3.82 & 2.66;~2.76;~2.98 \\ 
\hline
\end{tabular}
\caption{\label{tab:net_param} 
Parameters of the communities extracted from three analyzed time series of the QG3 model (the time series are referred to as TS1, TS2 and TS3). 
First row: contribution of a community to the network modularity. 
Second row: the number of SFA states (number of days) belonging to a community.
Third row: mean contribution of a state to the modularity of its community. 
}
\end{table*}


\subsection{Reanalysis data}

\begin{figure*}[ht!]
\includegraphics[width=\linewidth]{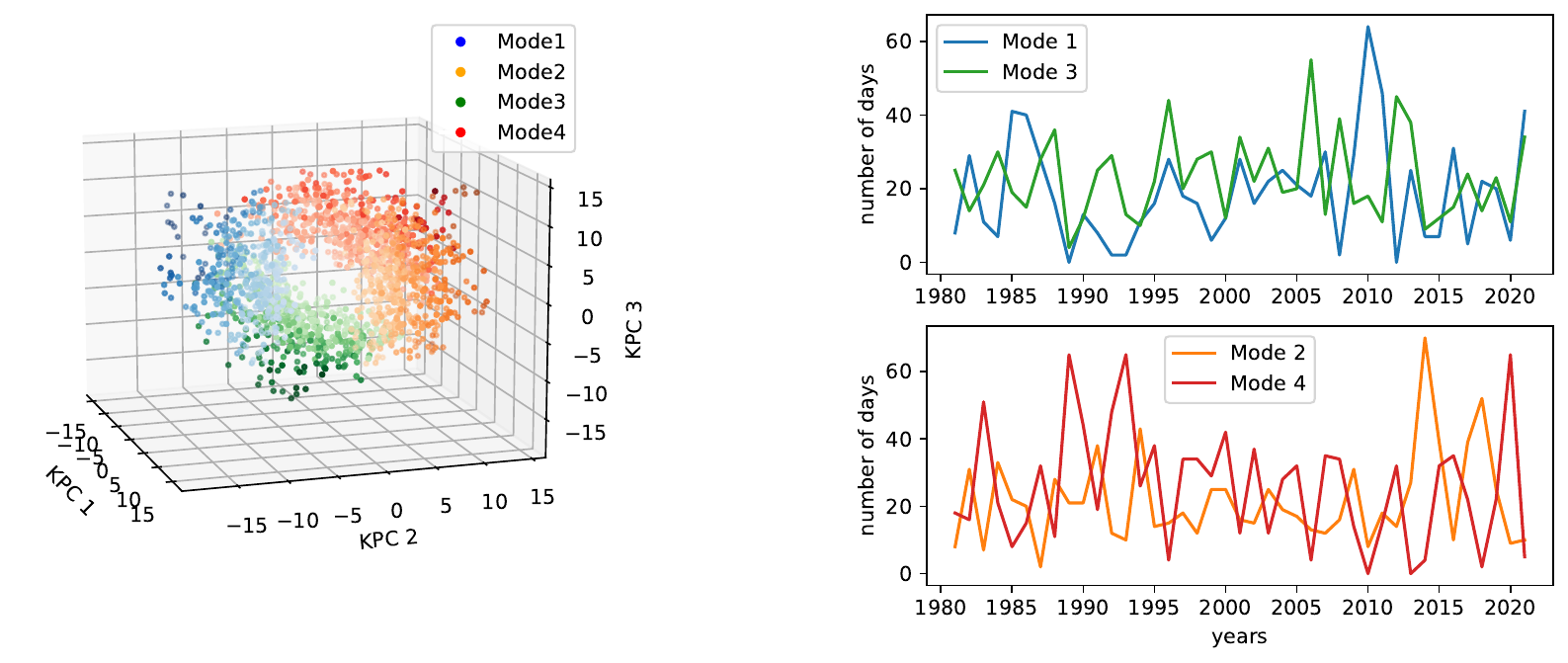}
\caption{Partitioning of HGT states into regimes. Left panel: 50\% of the most typical states of each regime (mode) in the space of three leading KPCs. Color saturation corresponds to the centrality value of a state in its community. Right panel: number of days per winter related to different regimes. The years of January of each winter are shown (e.g., 2001 corresponds to the winter 2000-2001).}
\label{fig:HGT-kpcs}
\end{figure*}

\begin{figure*}[ht!]
\includegraphics[width=\linewidth]{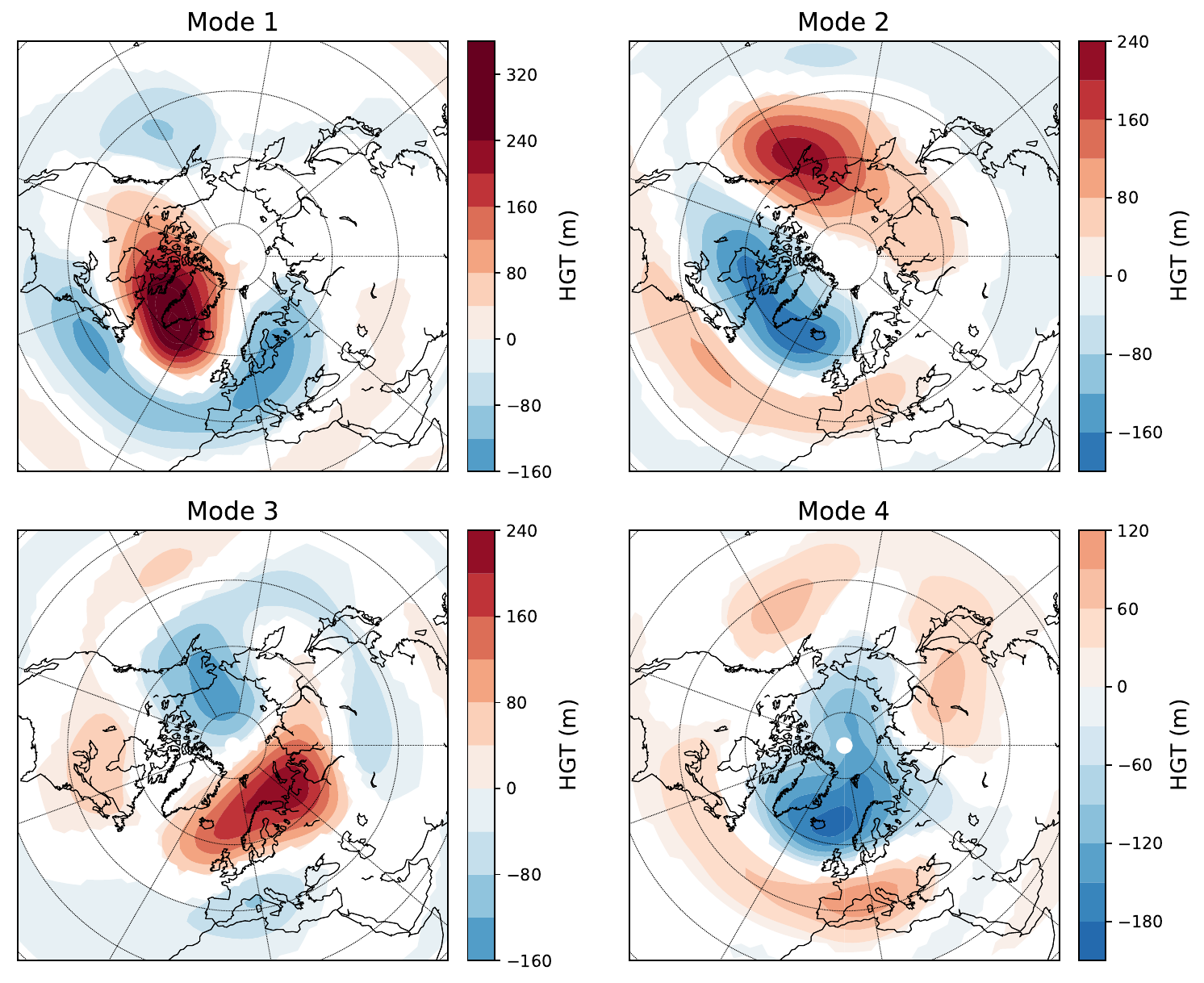}
\caption{Composite patterns of HGT corresponding to the obtained regimes. The composites are calculated as HGT averaged over 20\% most typical states of each of the regimes. { Only values that are significantly different from zero by Student's t-test with a critical value of 0.01 are shown.}}
\label{fig:comp-DJF}
\end{figure*}

\begin{figure*}[ht!]
\includegraphics[width=\linewidth]{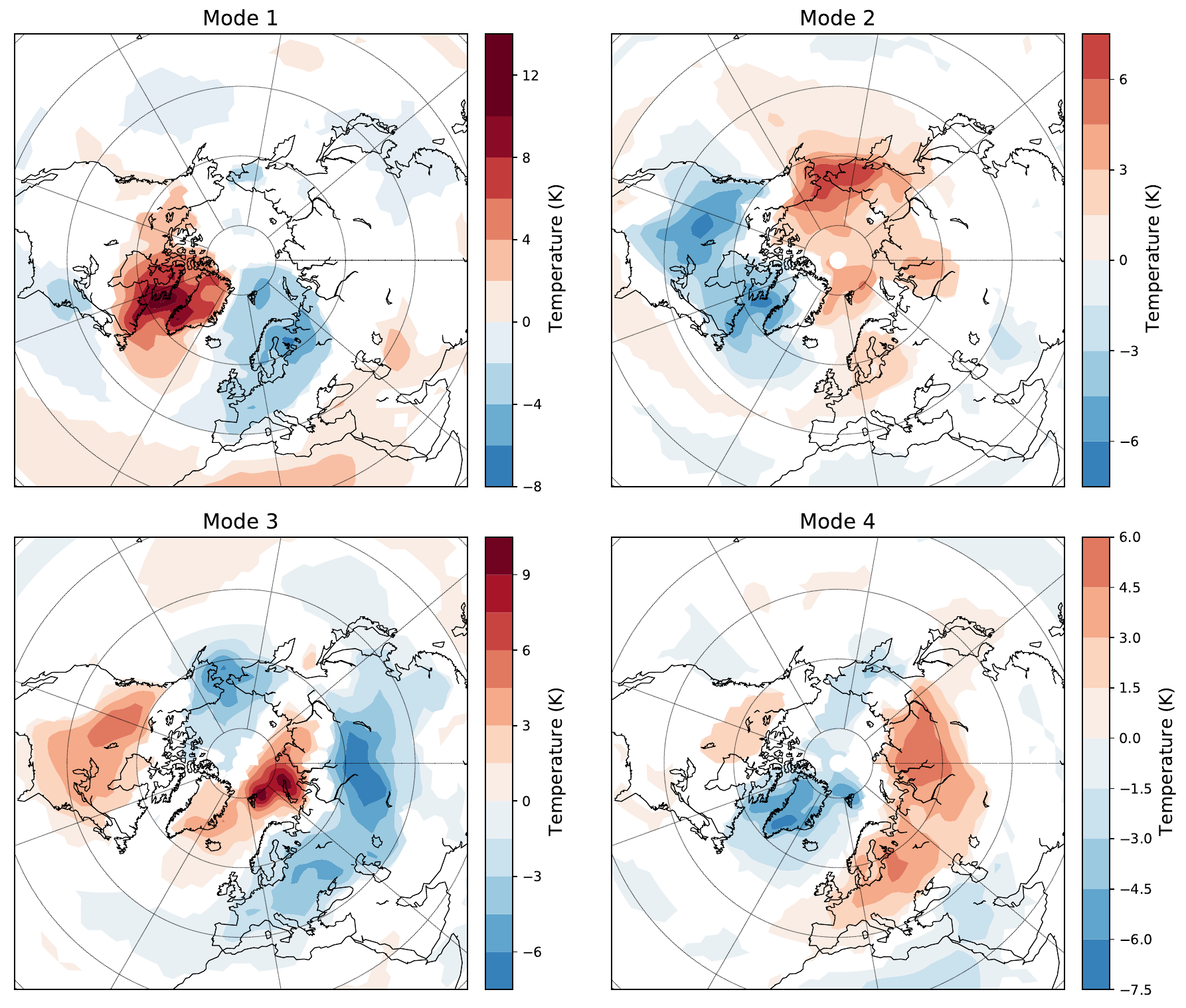}
\caption{Composite patterns of surface air temperatures (SAT) corresponding to the obtained HGT regimes. The composites are calculated as SAT averaged over 20\% most typical states of each of the regimes. { Only values that are significantly different from zero by Student's t-test with a critical value of 0.01 are shown.}}
\label{fig:comp-DJF-SAT}
\end{figure*}

For the HGT reanalysis data we detect four regimes (Fig.~\ref{fig:HGT-kpcs}, Tab.~\ref{tab:net_param_HGT}). 
These communities are embedded well in the space of 
three leading KPCs, organized into a loop/ring (Fig.~\ref{fig:HGT-kpcs}). The composites form typical HGT patterns in each community (Fig.~\ref{fig:comp-DJF}). Additionally, in order to study weather 
impacts of the detected circulation regimes, we take the same dates as used for the HGT composites to calculate the composites of 
near-surface air temperature anomalies\footnote{We use air temperature data at 0.995 sigma level taken from the NCEP/NCAR 
reanalysis \cite{Kalnay1996}. The anomalies were produced by deseasonalizing in the same way as with the HGT data (see 
Sec.~\ref{sec:data}.\ref{sec:data_rean}).} (Fig.~\ref{fig:comp-DJF-SAT}). 

The composites (Fig.~\ref{fig:comp-DJF}) demonstrate qualitatively different structures of atmospheric anomalies related to the 
detected regimes. The first regime is characterized by anticyclonic anomalies south of Greenland, projecting onto negative AO and 
NAO. This pattern blocks the transport of warm air from the Atlantic to Europe, while increasing the advection of subtropical air 
to the northeast of North America, and advecting warm/humid air into North Africa (Fig.~\ref{fig:comp-DJF-SAT}). The second 
regime features a positive NAO and PNA\cite{WallaceGutzler1981,Hannachi+2017}, manifested by a lowering pressure over the North Atlantic and increasing pressure in the 
northeastern Pacific Ocean. Anticyclonic anomalies in the Pacific Ocean block zonal airflow and lead to cooling in northern USA 
and Canada and warming in eastern Russia and Arctic. The third regime with a high pressure center over northwest Russia and 
Scandinavia slows down zonal air transport in the Euro-Atlantic region, leading to extremely cold winters in Europe and heating 
the Arctic Ocean area north of central Russia. Simultaneously, stable zonal flow over the north Pacific induces warm conditions 
in Canada (Fig.\ref{fig:comp-DJF-SAT}). Finally, the fourth flow pattern relates to positive AO and NAO\cite{Thompson1987,Wallace2000}; it is characterized by 
increased zonal airflow in the north Atlantic, providing warmer than normal winters in Europe and Russia (Fig.~\ref{fig:comp-DJF-SAT}).

The number of days in each winter corresponding to a given flow regime (Fig.~\ref{fig:HGT-kpcs}), 
as well as in the KPC time series (not shown), we observe pronounced 
strong inter-annual variability of dominating types of behavior of the winter atmospheric circulation.
In particular, we see that regimes 1 and 3, leading to cold weather in Europe, can be correlated on a large scale. 
The amplitude of 
the inter-annual variability is not regular; e.g., there are sets of extreme winters with strong domination of a single regime (see, for 
example, the abnormal winters 2009-10 and 2019-20 showing domination of regimes 1 and 4, respectively). 
Regimes 1 and 2 occur less often than 3 and 4 (Fig.~\ref{fig:HGT-kpcs}, Tab.~\ref{tab:net_param_HGT}), although they have 
larger modularity per state, i.e., the HGT states belonging to them are more distinguishable. 

\begin{figure}[ht!]
\includegraphics[width=\linewidth]{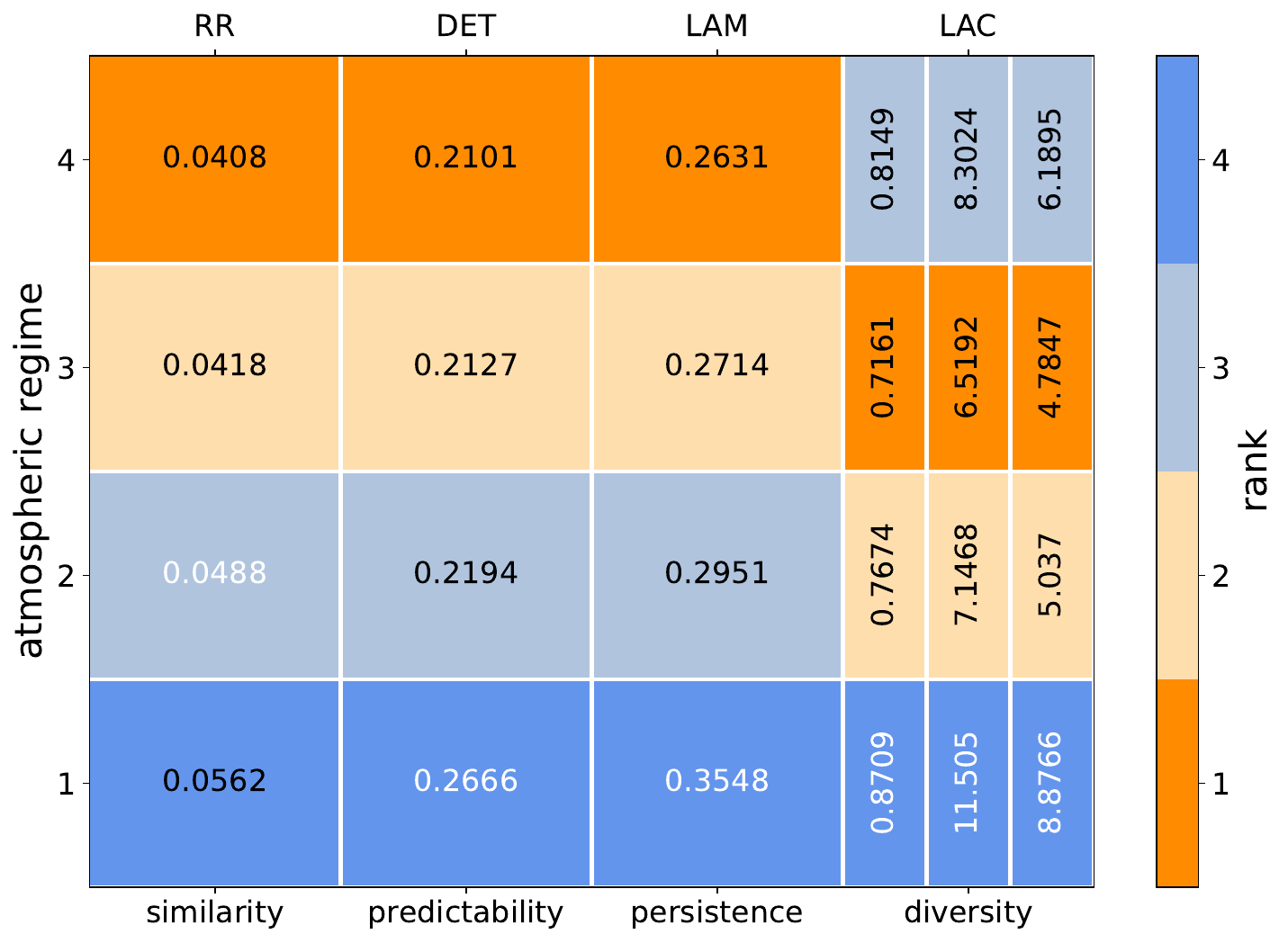}
\caption{Results of recurrence quantification analysis for the atmospheric regimes obtained in the reanalysis data set. The four different RQA measures are labelled with their respective interpretation. Values that significantly exceed the values corresponding to randomly obtained partitions (see the text) are printed in white; other values are printed in black. Color coding illustrates the ordering of the RQA values (ascending column-ranking) for better comparability.}
\label{fig:rean-RQA}
\end{figure}

Atmospheric patterns characteristic for the first regime are rendered persistent and predictable at 
{ intra-seasonal}
time scales 
(Fig.\ref{fig:rean-RQA}). However, this regime exhibits high 
diversity,
indicating strong variability at 
{ longer (inter-annual and decadal)}
time scales. This corroborates the general finding that atmospheric blocking structures entail stationary winter atmospheric circulation while their prediction at inter-annual to decadal time scales is cumbersome. We 
find significantly high similarity for the second regime, implying that the spatial anticyclonic anomaly patterns characteristic 
for this regime are comparable between different years. Note that high similarity obtained for the first and second regimes is 
likely the source of large values of the modularity per state within these regimes (see Tab.~\ref{tab:net_param_HGT}).
Conversely, we find that atmospheric conditions as identified in the third regime which, e.g., often result in extremely cold 
European winters are poorly predictable at 
{ intra-seasonal} 
time scales. Finally, positive AO and NAO phases as represented by the 
fourth regime are dynamically opposite to the atmospheric blocking structures (negative AO and NAO) captured by the first regime.

{ This is in agreement with Woollings et al.\cite{Woollings+2010a}, and references therein, 
who investigated the NAO time scales. They found that the two phases of the 
NAO have intrinsically different decay characteristics, with the negative NAO 
events showing enhanced persistence associated with blocking, see also Ref. \onlinecite{Onskog+2020}. 
Woollings et al.\cite{Woollings+2010b} investigate the jet positions over the North Atlantic region. They found that the southern jet position, associated with the negative NAO phase (and Greenland blocking) has smaller tendency, and therefore  more persistent (see Ref. \onlinecite{Hannachi+2017})  than the northern position, associated with the positive NAO phase.}

\begin{table*}[ht]
\centering
\begin{tabular}{|l|l|l|l|l|}
\hline
~ & Community 1 & Community  2 & Community 3 & Community 4\\
\hline
Modularity  & 0.109 & 0.122 & 0.115 & 0.13\\
\hline
Number of states & 777 & 890 & 940 & 1062\\
\hline
Modularity per state, ($\cdot10^{-4}$) & 1.4 & 1.37 & 1.22 & 1.22\\ 
\hline
\end{tabular}
\caption{\label{tab:net_param_HGT} 
The same as in Table \ref{tab:net_param}, but for the regimes obtained from the reanalysis HGT data set. 
}
\end{table*}

\section{Summary and conclusions}

The proposed method allows us to (1) reveal recurrent regimes of atmospheric circulation from spatially distributed observations, 
and, simultaneously, (2) obtain a set of dynamical variables serving as an embedding for the regimes. A combination of the two 
nonlinear data-driven approaches – KPCA and recurrence analysis – provides comprehensive investigation of the mode content of the 
observed dynamics, including the regime identification, their dynamical representation and characteristics, and analyzing the 
dynamical properties of inter-regime evolution.  Both parts of the method are based on constructing the same kernel matrix that 
consists of pairwise similarities between atmospheric states at different dates. In the first part this matrix produces the 
recurrence network, which when partitioned yields separation of all observed states into the regimes. RQA applied to the obtained 
submatrices relating to different regimes (or communities), helps to study important properties of temporal evolution of the 
regimes, e.g., predictability, persistence, similarity, and intermittency. In the second part, the principal components of the 
kernel matrix (KPCs) are used to construct a space in which the states belonging to different regimes are well-separated.     

{ It is worth noting that the KPCA with Gaussian kernels is very close to the diffusion map method \cite{Coifman2006}, based on decomposition of a diffusion operator reconstructed from the data. More precisely, the 1-step diffusion maps should give the same, up to a transformation, basis of principal components as Gaussian KPCA. Thus, we can expect a clear separation of the regimes in the diffusion space too. Moreover, probably the use of n-step diffusion map space as an embedding for the regimes may be more effective, since it is more robust to noise due to many paths between network’s nodes being involved. In this work we demonstrate that even the basic Gaussian KPCA represents well the recurrence network communities due to the same distance matrix used in both the RP and Gaussian kernels. In future the diffusion maps can be adopted for this purpose.}

We demonstrate, using both model and observation data, that the detected regimes of the Northern Hemisphere mid-latitude 
winter atmosphere correspond to qualitatively different states, which cover the well-known modes NAO, AO, and PNA.  We show 
that typically only a few leading KPCs are sufficient for the embedding of the regimes. Thus, these KPCs can be used as 
dynamical variables describing the alternation of the obtained regimes, and future works can aim at predictive data-driven models 
of their dynamics (see, e.g., Refs. \onlinecite{Gavrilov2019,Mukhin2019}). Moreover, having the dynamical variables representing the 
atmospheric modes, we can state a problem of finding long-term climatic predictors (e.g., ENSO, QBO, solar cycle, etc.) making it 
possible to elaborate a scheme for inter-annual forecast of dominating weather patterns.  

\begin{figure*}
\includegraphics[width=0.9\linewidth]{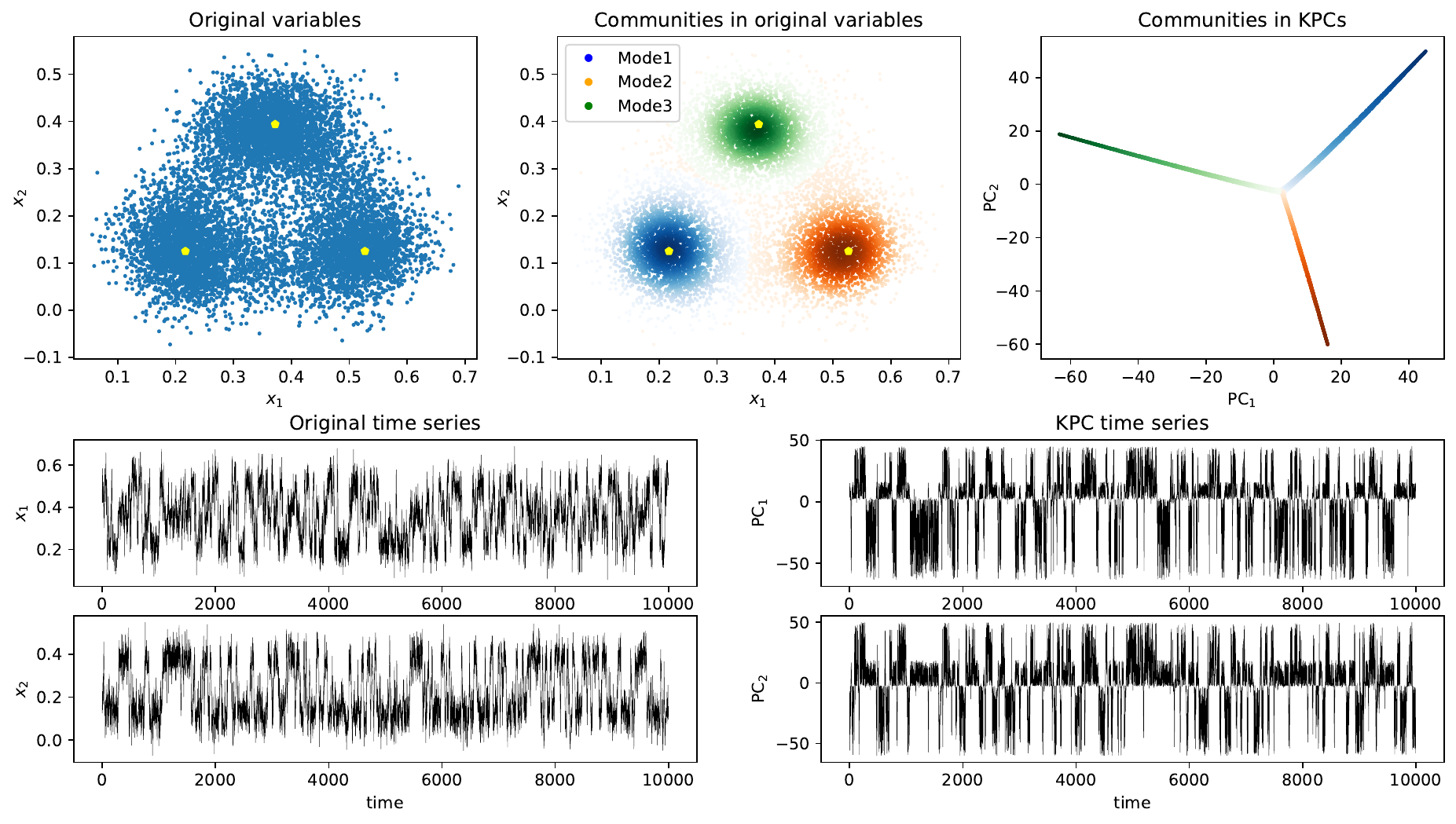}
\caption{Analysis of time series of the three-well model (Eq. \ref{eq:3well}-\ref{eq:profile}) based on the simple Euclidean distance between state vectors. Upper panels (from left to right): analyzed time series presented in original variables; the same, but with states marked according with the regimes they belong to; analyzed time series presented on the plain of leading two KPCs. Yellow dots mark ``theoretical'' centers of the potential function $V(\cdot)$. Color intensity corresponds to centrality of a state in its community. Low panels: time series of original variables (left) and leading two KPCs (right).}
\label{fig:3well-nonorm}
\end{figure*}

\begin{figure*}[ht!]
\includegraphics[width=0.9\linewidth]{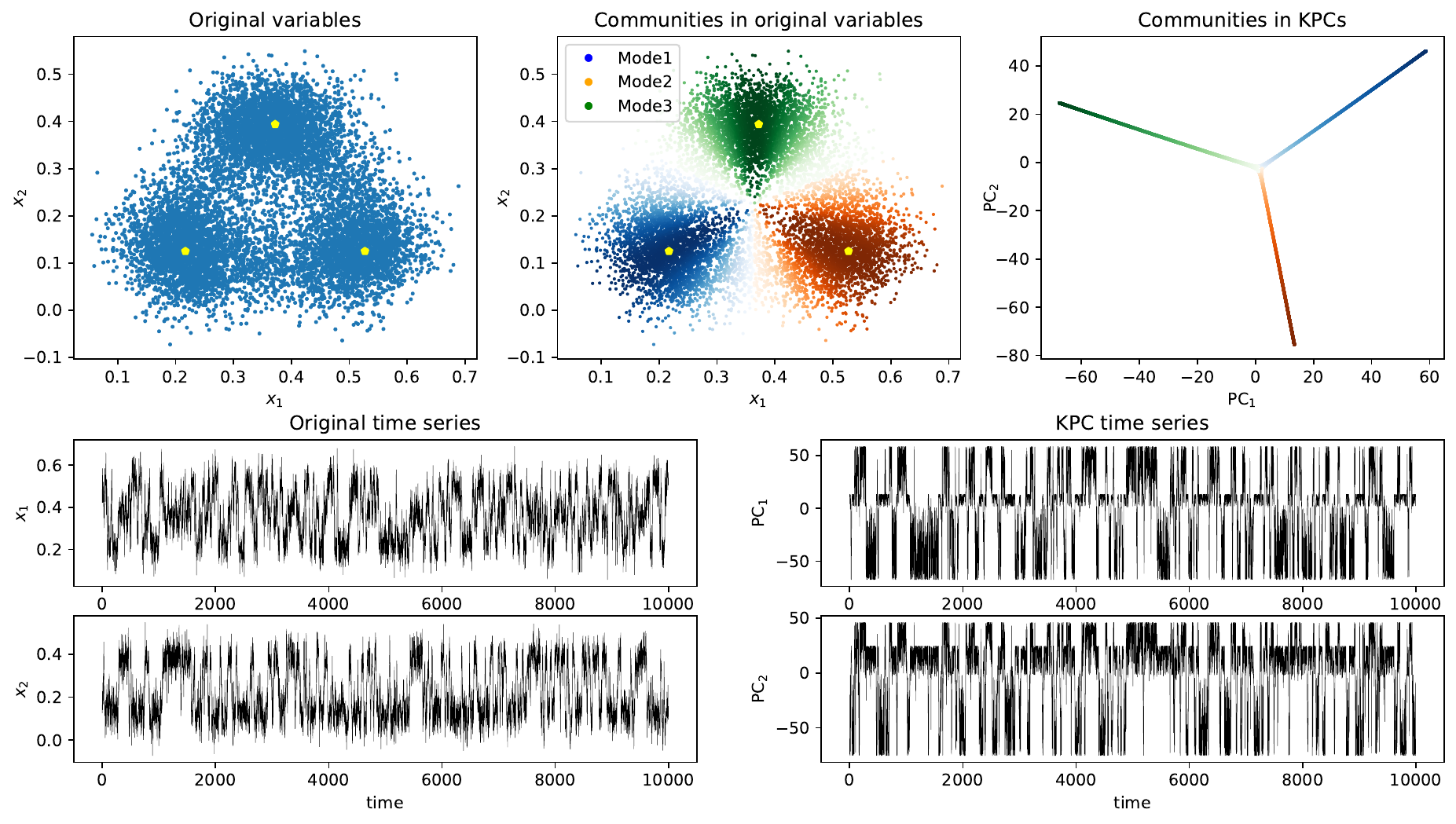}
\caption{The same analysis as in Fig. \ref{fig:3well-nonorm}, but based on Euclidean distance between normalized vectors defined by Eq. \ref{distance}.}
\label{fig:3well-norm}
\end{figure*}

The kernels Eq.~(\ref{kernel}), based on the distance Eq.(\ref{distance}), reflect similarity between two short-term 
patterns, each realized within one day. This leads to extracting the recurrent but not necessary persistent  patterns (the 
persistence is separately studied by the RQA), which are distributed over the whole mid-latitude belt. However, we can change 
the distance definition, adapting it to the desired properties of regimes to extract. For example, we can target  the method to 
long-living (persistent) regimes via time-lag extension of states, or use other weightings, Eq.~(\ref{weight}), to emphasize some 
geographical regions. Those topics are left for future research.



\bibliography{references}

\section*{Supplementary material}
See the supplementary material that includes
supplementary figures 1 and 2. 

\begin{acknowledgments}
{ The authors are grateful to the two anonymous reviewers for fruitful suggestions and comments. They helped to improve the article significantly.
The QG3 model code is provided by the courtesy of Dmitri Kondrashov (UCLA) and Fabio d’Andrea (ENS).}
This work was supported by the grant \#22-12-00388 of the Russian Science Foundation (D. Mukhin) and Deutsche Forschungsgemeinschaft in the context of the DFG project MA4759/11-1 ‘Nonlinear empirical mode analysis of complex systems: Development of general approach and application in climate’ (T. Braun).
\end{acknowledgments}

\section*{Author Declaration}
\subsection*{Conflict of Interest}
The authors have no conflicts to disclose.
\subsection*{Author Contributions}
DM and AH developed the method for detecting regimes, TB and NM analyzed the regimes by the RQA approach. All the authors analyzed results and contributed to writing the manuscript. 

\section*{Data Availability Statement}
The data that support the findings of this study are openly available in NCEP/NCAR reanalysis\cite{Kalnay1996} archive at
\url{https://psl.noaa.gov/data/gridded/data.ncep.reanalysis.html}. { The time series of the QG3 model are available upon request.}

\appendix

\section{Simple model example}

{ This section demonstrates the ability of the proposed method to detect and separate recurrent states in a simple situation, when the true solution is known. We use a toy two-dimensional stochastic model with three-well potential, which was suggested in the work \onlinecite{Hannachi2001} for testing a clustering method and then also used in the work \onlinecite{Kravtsov2005}:
\begin{equation}\label{eq:3well}
\begin{array}
dd{\bf X}=-\nabla V({\bf X})+\epsilon dW_t \\
~\\
V({\bf X})=u({\bf X-A}_1)+u({\bf X-A}_2)+u({\bf X-A}_3)\\
+b\left|{\bf X-A}_c\right|^2,
\end{array}
\end{equation}
where $W_t$ is the Wiener process, points $A_1=(0,0)$, $A_2=(2a,0)$ and $A_3=(a,a\sqrt{3})$ are the centers of profiles $u(\cdot)$ which takes the form
\begin{equation}\label{eq:profile}
u({\bf X})=-\alpha\exp{\left[\frac{1}{|{\bf X}|^2-a^2}\right]}.
\end{equation}
The potential function $V({\bf X})$ has three minima placed in the vertices of a isosceles triangle centered in $A_c=\left(a,\frac{a}{\sqrt{3}}\right)$. Here we use the same values of parameters as in Ref. \onlinecite{Hannachi2001}: $\alpha=21$, $a=0,87$, $b=0,12$ and $\epsilon=0.05$.
This model produces random walks around the local minima of the potential function, provided that the phase trajectory spends more time near the centers of the potential than anywhere else. Thereby, the model simulates the situation with three regimes, or recurrent states in the phase space.

We integrate this model by Euler method with the time step 0.01. To exclude from consideration the points which are close to each other due to the temporal ordering, we take only each 100th point during the integration. For analysis we use 10,000-point time series (Fig. \ref{fig:3well-nonorm} and \ref{fig:3well-norm}) taken after 20,000-step spin-up period. 

We applied the presented method to this data, based on two different distances: (i) simple Euclidean distance $d_{ij}=\left|{\bf X}_i-{\bf X}_j\right|$ and (ii) Euclidean distance between normalized vectors given by Eq. \ref{distance}. In both cases we detect three regimes, central points of which are separated well in the plane of leading two KPCs (see Fig. \ref{fig:3well-nonorm} and \ref{fig:3well-norm}). In the projections of the phase trajectory to this plane, states that belong to different regimes fall on different linear manifolds. At the same time, the time series of KPCs describe transitions between areas of the regimes. However, distribution of the centralities substantially depends on the distance choice: if in the first case it emphasizes the maxima of the probability density in 2d-space, in the second case it captures rather the density of angles of centered states vectors, because the normalization in Eq. \ref{distance} attaches all vectors to the unit sphere. The latter is especially important in climate applications above, where we are interested in capturing the shapes of spatial patterns rather than their amplitudes.}



\begin{figure*}[ht!]
\centering
\includegraphics[width=.5\linewidth]{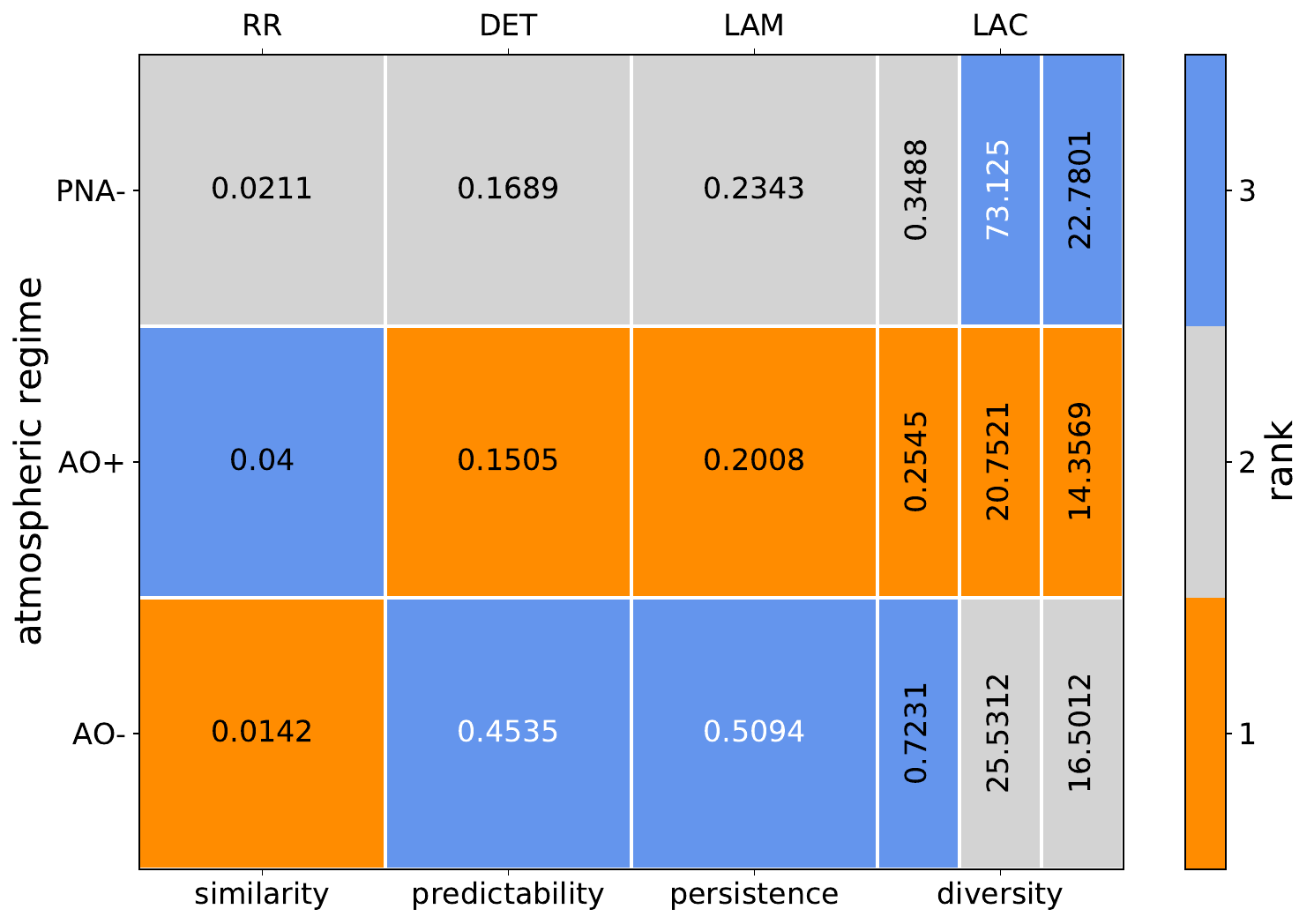}\\
\includegraphics[width=.5\linewidth]{QG3_RQA2.pdf}\\
\includegraphics[width=.5\linewidth]{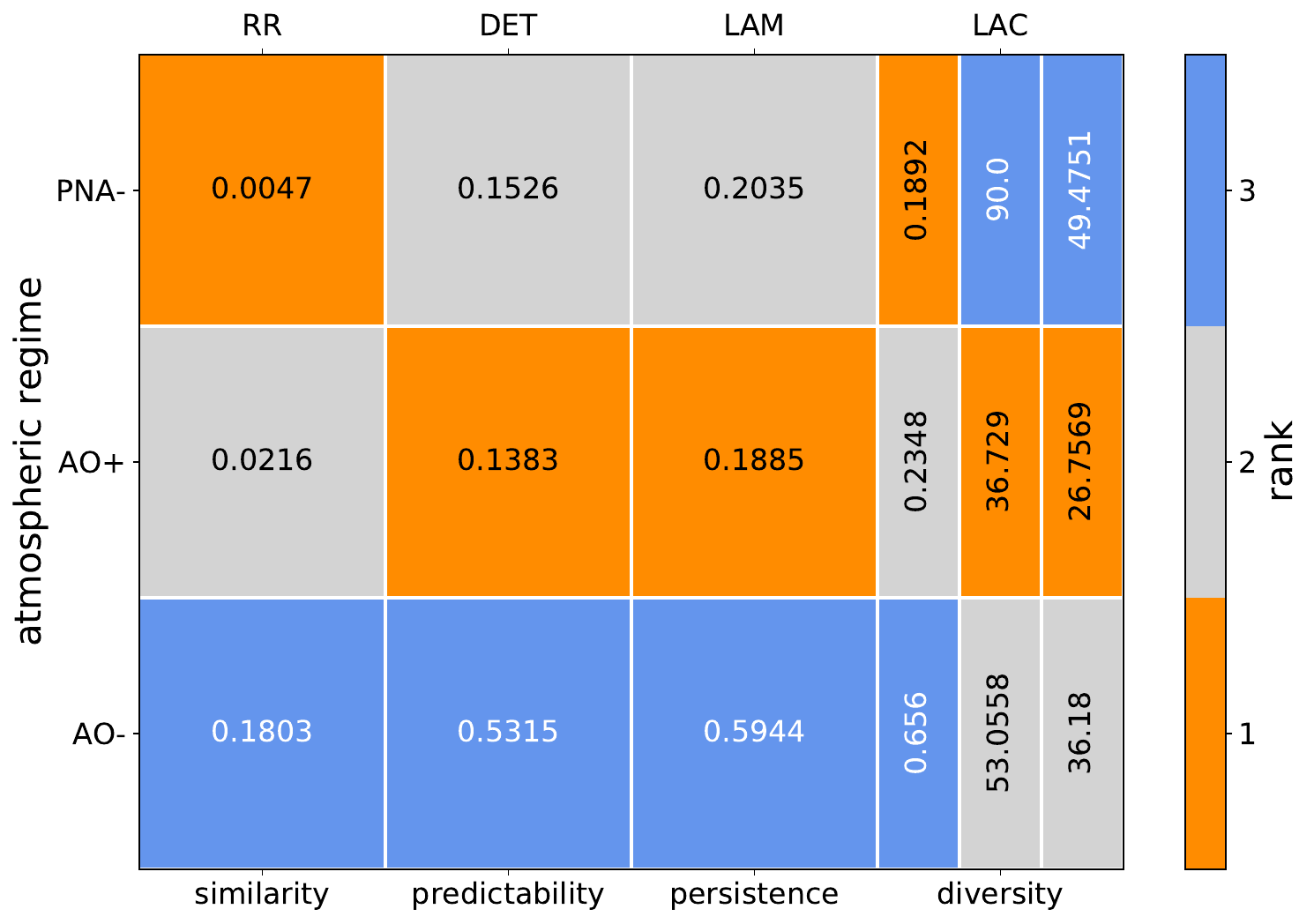}
\caption*{
{\bf SUPPL. FIG. 1.}
Comparison of results of recurrence quantification analysis for the atmospheric regimes obtained in the QG3 model data set between the three different time series. The four different RQA measures are labelled with their respective interpretation. Significant (insignificant) values are printed in white (black). Color coding illustrates the ordering of the RQA values (ascending column-ranking) for better comparability.}
\label{fig_suppl:rqa_qg3}
\end{figure*}

\begin{figure*}[ht!]
\includegraphics[width=\linewidth]{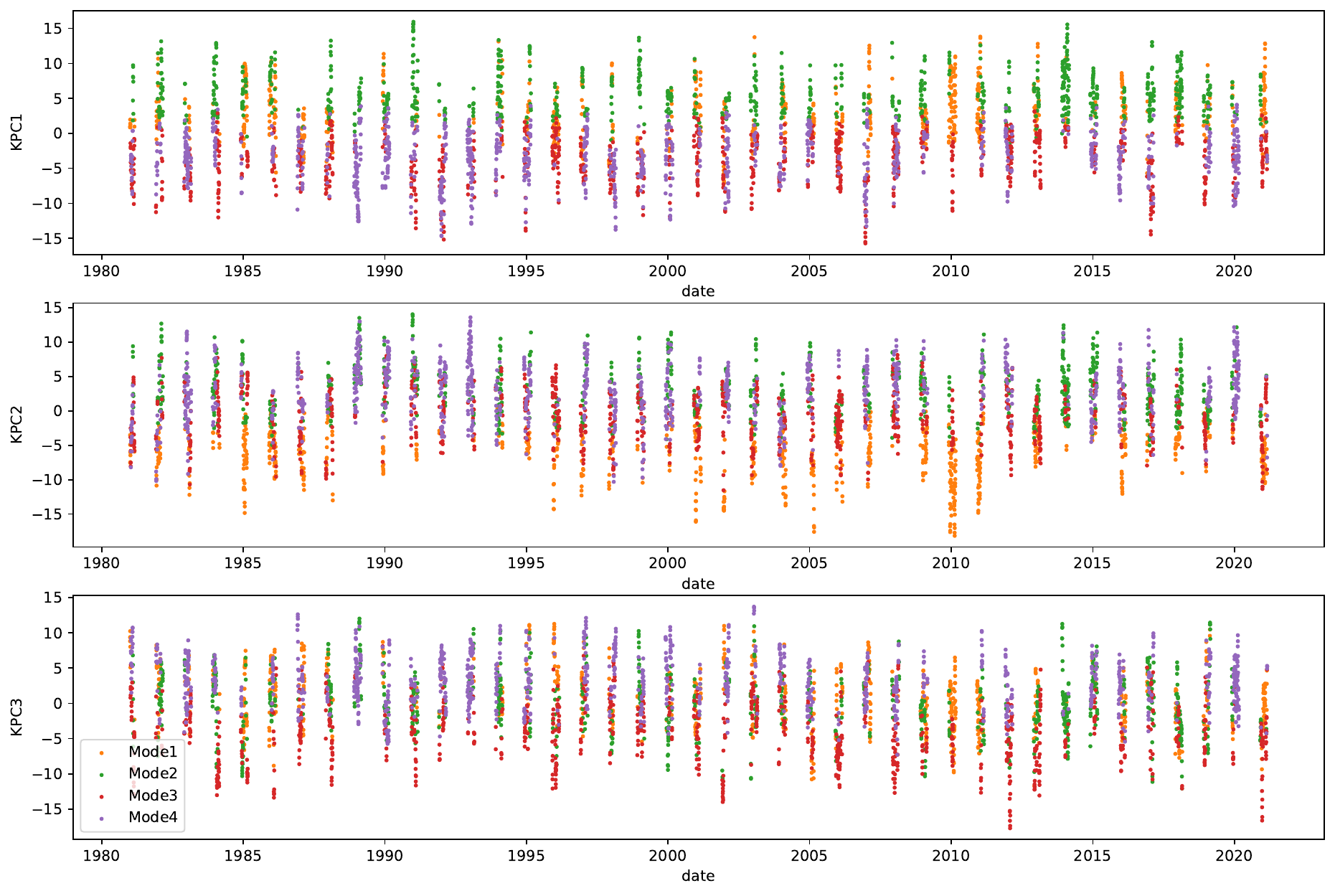}
\caption*{{\bf SUPPL. FIG. 2.}
{ Time series of the leading three KPCs of the reanalysis data. Different colors correspond to the obtained regimes. Since only the winter months are considered, the time series are split into segments separated by equidistant gaps.}
}\label{fig_suppl:rp_rean}
\end{figure*}
\begin{figure*}[ht!]
\includegraphics[width=\linewidth]{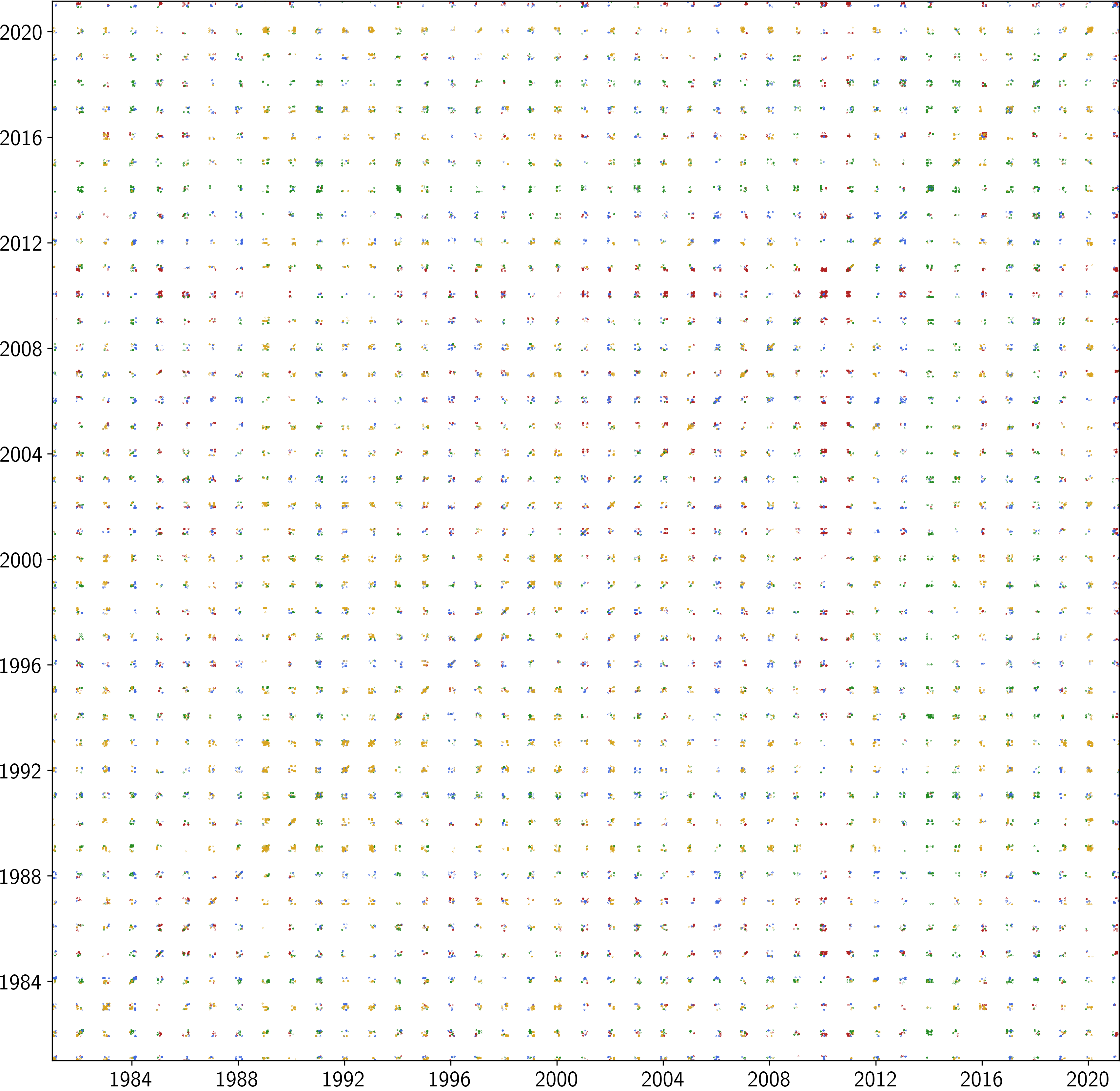}
\caption*{{\bf SUPPL. FIG. 3.}
Recurrence plot of atmospheric patterns obtained from the reanalysis data. Recurrences within a given regime are color coded accordingly. Since only the winter months are considered, the RP is split into sub-RPs separated by equidistant gaps.}\label{fig_suppl:rp_rean}
\end{figure*}

\end{document}